\documentclass[11pt,thmsa,a4paper]{article}
\RequirePackage[OT1]{fontenc} \RequirePackage{graphicx}
\RequirePackage{amsthm} \RequirePackage[cmex10]{amsmath}
\RequirePackage{natbib}
\RequirePackage[colorlinks,citecolor=blue,urlcolor=blue]{hyperref}

\usepackage{amssymb,amstext}
\usepackage{booktabs,graphics,fancybox,ulem,color,pifont,amsfonts,amsbsy}
\usepackage{subfig,epsfig,calc}
\usepackage{eurosym,picture,tikz,amsthm,bm}
\usepackage[dvipdfm]{geometry}
\usepackage{verbatim}
\usepackage{rotating}

\usepackage{graphicx}
\begin{document}

\title{A Varying Coefficient Model for Assessing the Returns to Growth To Account for Poverty and Inequality} 

\author{Max K\"ohler \footnote{Max K\"ohler, University of G\"ottingen, CRC Poverty, Equity and Growth, G\"ottingen, Germany},
Stefan Sperlich\footnote{Universit\'e de Gen\'eve, Geneva School of Economics and Management, Switzerland, stefan.sperlich@unige.ch},
Jisu Yoon \footnote{University of G\"ottingen, Faculty of Economics, G\"ottingen, Germany}}

\date{DP version 2016}

\maketitle

\begin{abstract} \noindent
Various papers demonstrate the importance of inequality, poverty and the size of the middle class for economic growth. When explaining why these measures of the income distribution are added to the growth regression, it is often mentioned that poor people behave different which may translate to the economy as a whole. However, simply adding explanatory variables does not reflect this behavior. By a varying coefficient model we show that the returns to growth differ a lot depending on poverty and inequality. Furthermore, we investigate how these returns differ for the poorer and for the richer part of the societies. We argue that the differences in the coefficients impede, on the one hand, that the means coefficients are informative, and, on the other hand, challenge the credibility of the economic interpretation. In short, we show that, when estimating mean coefficients without accounting for poverty and inequality, the estimation is likely to suffer from a serious endogeneity bias.
\end{abstract}

\newpage

\section{Introduction} \label{sec:intro}

The literature shows that the variables inequality, poverty and the size of the middle class are important for economic growth. Usually the authors add the variables addressed in their special question to the growth regression and observe the effects of these variables. When explaining why certain measures of the income distribution are added, it is often explained that poor people behave differently from rich people and therefore, the economies as whole behave differently according to their level of poverty, inequality and the size of their middle class. However, adding explanatory variables does not reflect this behavior. For example, it is hard to believe that a poor economy, in which a high number of poor people cannot provide collateral and therefore, do not have access to the credit market, has the same returns to investments in physical capital as a richer economy. This paper empirically investigates the effects of measures of the income distribution on the coefficients of the drivers for economic growth. More precisely, we consider three parameters of the income distribution, namely poverty, inequality and the size of the middle class and investigate the influence of each of these variables on the coefficients of the drivers for growth proposed by \citet{mankiw_1992}. For this purpose, we formulate and apply a varying coefficient model. We focus on a panel data analysis using data from several countries over time.

The literature on growth, poverty and inequality is much too abundant to briefly discuss it here, but the explicit modeling of the impact of the latter on growth is not that frequent; see
\citet{bourguignon_2004} for a review.
Generally, while the literature basically agrees on the fact that inequality affects economic growth, there is no consensus about the size or the direction of the effects. Even when concentrating on the most recent contributions you find studies arguing for the inverted U-shaped relationship \citep[e.g.,][]{tapia_2015}, positive effects of inequality on growth \citep[e.g.,][]{forbes_2000} and negative effects \citep[e.g.,][]{malinen_2013}. Some argue that inequality has different effects on growth depending on the context \citep[e.g.,][]{halter_2014}. 
On the other hand, there is a consensus that poverty affects growth negatively \citep[e.g.,][]{ghatak_2015}, while the size of the middle class is considered to be beneficial to economic growth \citep[e.g., Aristotle 306 BC, ][]{chun_2016,easterly_2001}. 

The channels through which poverty, inequality and the middle class influencing economic growth are diverse. This suggests that economies behave differently according to these variables. When credit market works imperfectly, lenders will demand collaterals to cope with asymmetric information. Since the poor cannot provide them, then their investment plans are likely to be wasted. Consequently, poverty and a high level of inequality could lead to low economic growth \citep{banerjee_1993,perry_2006}. Furthermore, high inequality and a small middle class may deter economic growth by higher socio-economic instability \citep{alesina_1996} and conflicts  \citep{alesina_1994b,rodrik_1998}. Poverty also influences economic growth negatively through the financial market development \citep{ravallion_2010} and too much risk averse behavior of the poor \citep{banerjee_2000}. \citet{sachs_2004} and \citet{mesnard_2006} support the existence of a poverty trap which has obviously negative implications for economic growth.

Since economies behave differently according to poverty, inequality and the middle class, the mean returns to the drivers of economic growth may not be informative. Imagine that the population (and sample) can be divided into two groups, on composed by poor countries, and the other by rich ones, 
and consider the simplified growth regression 
\begin{equation} \label{ex-1}
 growth = \alpha + \beta \cdot (growth \ driver) + error \ . 
\end{equation}
In this situation it is very likely to hold that $\beta_{poor} \neq \beta_{rich}$
when studying the two groups separately. However, when pooling them, then the mean coefficient 
does not only reflect a theoretical parameter that is little helpful. 
Furthermore, there are problems when bringing the model to data. Poor countries have systematically weaker databases and therefore, the estimation of such a $\beta_{mean}$ is highly suspicious to suffer from a sample selection bias. 
Moreover, as the deviations from the mean coefficient are highly suspicious to move simultaneously with either the covariates, the dependent variable or both, you face an endogeneity problem in the sense that this mean-parameter cannot be estimated from model (\ref{ex-1}) with a standard OLS 
applied to the merged sample. 

Certainly, separating the coefficient into $\beta_{poor}$ and $\beta_{rich}$ from the beginning
could solve these problems. In practice however, we have many different growth drivers and a set of countries that cannot simply be separated into the groups $poor$ and $rich$. It is obvious, that then the non-modeling of $\beta$ becomes a serious endogeneity problem which typically 
cannot be solved, even not with more sophisticated IV methods.
This motivates to estimate the growth regression with a varying coefficient model in which a ``continuous transition" from poor to rich is possible. This transition is explained by the country's individual levels of poverty, inequality and middle class in each year.

There exist also many other reasons to use a varying coefficient model for growth regression. 
While many authors include additional covariates like poverty and inequality additively to the growth regression, one could equally well argue that it is hard to understand why these should be 
essential production factors of GDP on their own; recall that the classical growth regression model
is derived from the Cobb Douglas specification of a production function applied to the countries DGP. The arguments brought up in the literature why and how poverty and inequality affect growth
insinuate that these covariates have an impact on the efficiency and thus the return of the production factors rather than on GDP directly. This suggests that the coefficients of
the classical growth model should be modeled accordingly; it does not suggest to add
these covariates additively to the model.

This article is organized as follows. Section \eqref{sec2} first revisits the augmented Solow growth model, deals with collecting reasonable measures of the covariates, and then discusses different estimation issues. The section concludes with formulating the desired varying coefficient model
for growth regression.
The results of estimation are given in section \eqref{sec:results}. We first present the results 
for a model where the mean growth per worker is the dependent variable. It will be seen that the 
coefficients change dramatically over poverty, inequality and middle size. This motivates another question which is related to the pro-poor growth discussion. If our derivatives of the income distribution affect the growth behavior of the economy, then it could be interesting to see
how do the poorer and the richer parts of the countries regarding growth? 
We do this in an admittedly quite simple way by just looking at a growth
regression with the dependent variable being either the average GDP of the lower income group 
(0 to 20\% quantile of the income distribution) and the richest (80 to 100\%  quantile of the income distribution) respectively. The resulting coefficients differ dramatically. 
Section \eqref{sec:conlcusion} concludes.

\section{Statistical Modeling and Data Collection}
\label{sec2}

\subsection{Growth Regression Model and Data}
\label{subsection:data}

We use the classic economic model for growth regression which explains the GDP production
with a Cobb Douglas function specification. After some algebra one has for country $i$ in year 
$t$
\begin{equation} \label{regression}
y_{it}=\rho y_{i(t-l)} + x_{it}'\beta + \eta_i + \nu_{it}, \ t=1+l,\ldots ,T, i=1,\ldots ,n,
\end{equation}
where $y_{it}$ is the logarithm of the per worker or capita GDP, 
$l$ some lag depth, vector $x_{it}$ contains the 
log of production factors (in the here considered augmented Solow-model, the physical and the human capital) and depreciation rates, $\eta_i$ country fixed effects, and $\nu_{it}$ a stochastic deviation with expectation zero. Sometimes time fixed effects are included but get 
insignificant when applying the Hodrick-Prescott filter \citep{hodrick_1997} to $y_{it}$ in order to 
get rid of business cycle effects.\footnote{This was also done for the time series of investment shares indicating the physical capital.} 

The data are based on Penn World Table 6.3 \citep[PWT,][]{heston_2009}, the World Development indicators \citep{worldbank_2009} and \citet{barro_2010}. The observations are obtained yearly from 1960 to 2007 for 81 countries. The depreciation rate is the sum of the depreciation rate of capital, the growth rate of productivity and the population growth \citep{mankiw_1992}. The sum of the depreciation rate of capital and the growth rate of productivity is approximated by 5\% per year for all countries, which is added to the population growth. The logarithm of the depreciation rate is denoted by $lnn_{it}$. The saving rate of the economy is approximated by the relative investment share of the real GDP, which is denoted by $lnsk_{it}$. The human capital is calculated based on the educational attainment data from \citet{barro_2010}. Since the raw data are available every five years, interpolation splines is used to impute the missing values. We denote the logarithm of the yearly educational attainment by $lnattain_{it}$. Based on the country rating system from \citet{heston_2009}, only countries with sample quality between A and C are kept, dropping D grading samples. Furthermore, only complete time-series are incorporated for the relevant variables. This also excludes countries that were separated in a sub-period, for example Germany. 
Recall from our discussion above that the varying coefficient model will essentially 
alleviate (or even solve) the potential selection bias problem. 

We are going to investigate how the factor returns $\beta$ change along the
different income distributions. You could write
$$ y_{it}=\rho y_{i(t-l)} + x_{it}' \beta_{it} + \eta_i + \nu_{it}  \ , $$
where $\beta_{it}$ is (partly) explained by the income distribution in country $i$ at time $t$.
Since estimating $\beta_{it}$ as a function of densities would be far to complex,
we model some of the returns as functions of poverty, inequality and middle class, instead. 
Measures for these three variables are derived from the income distribution data from \citet{sala-i-martin_2006}. His data consists of hundred data points of income distribution for 138 countries every year from 1970 to 2000, and is based on the microeconomic income surveys from \citet{deininger_1996}, while the final values are imputed, adjusted and smoothed. We use the fraction of the total population with income less than one dollar (1999 price) per day as our measurement of poverty, which is denoted by $pov1d_{it}$. 
The Gini coefficient is our measurement of inequality, which is denoted by $gini_{it}$. We repeated the study with the Theil index obtaining the same results. The middle class is the share of the total income that the middle sixty per cent of the population earn \citep[cf.,][]{easterly_2001}. 
More precisely, let $g_{it}( prop )$ be the cumulative income as a function of the proportion $prop$ 
of the total population ($pop_{it}$) of country $i$ at year $t$, i.e.\ the Lorenz curve times total 
income. Then the relative middle class is 
$ middleclass_{it}=(g_{it}(0.8)-g_{it}(0.2))/g_{it}(1.0) $.
Furthermore, as indicated at the end of Section \ref{sec:intro}, we are also interested in knowing the average incomes of the richest and the poorest twenty per cent of each country in each year. 
The log average GDP per capita of the poorest twenty per cent of country $i$ at year $t$ is  
$y_{poor,it}=ln(g_{it}(0.2)/0.2pop_{it})$. 
This measure was introduced by \citet{ravallion_2003} for measuring pro-poor growth. In the same way, the log average GDP per capita of the richest twenty per cent of country $i$ at year $t$ is $y_{rich,it}=ln((g_{it}(1.0)-g_{it}(0.8))/0.2pop_{it}).$ We multiply the consumer price index \citep{worldbank_2009} appropriately to these variables, so that the reference year becomes 2005 coherent to the PWT. As these series might be subject of business cycles we smooth them, too
\citep[for details cf.][]{koehler_2011}. 

When combining all data, we end up with complete yearly time-series from 1970 to 2000 for the 81 countries. For our estimation we rely on the fixed effect model in the following as random effect model is unattractive given that the individual deviation $\eta_i$ is likely to be dependent from other determinants of economic growth. But as we face a dynamic panel model the estimates could suffer from the Nickell bias. Therefore we have calculated the size of the bias using the formulae from \citet{phillips_2007} for various large ($>0.9$) values of $\rho$. It turns out that only for the estimates of $\rho$ the bias is larger than $3\cdot 10^{-4}$. Considering the long time series of the data, this result is not surprising. Note that we calculated the bias for equation \eqref{regression}; adding more regressors will reduce the bias further \citep[see,][]{phillips_2007}. The Difference and the System GMM are inappropriate as $\rho $ is $>0.9$, a situation in which the former GMM performs especially bad and the latter most likely violates the necessary conditions. Alos, both GMM would suffer from too many instruments in our context, \citep[see e.g.][]{roodman_2009}.

\subsection{A Varying Coefficient Growth Model}
\label{subsection:vcm}

As discussed, we introduce a varying coefficient model for growth regression, in which the coefficients depend on poverty, inequality and the size of the middle class, summarized in variable $z_{it}$. To the best of our knowledge the use of varying coefficient models in econometrics started in the seventies; for example \citet{wachter_1970} for studying wage equations, and \citet{singh_1976} allowing for
coefficients with time-trends. A more generalized version, in which some of the coefficients are functions of other exogenous variables, is introduced in \citet{amemiya_1978}.
Nonparametric extensions are still quite recent \citep[for a review see][]{PBMLL2013}.
Finally, \citet{RodPooSoberon2015} consider semiparametric varying coefficient models 
for panel data in which the drivers of the coefficients vary over both dimensions, 
time and subjects. 
That is, as there is no reason to assume that the $\beta$-coefficients are constant either over time or countries, let us rewrite (\ref{regression}) as
\begin{equation}\label{regression_vc}
y_{it}=\rho y_{i(t-l)} + x_{it}' \beta_{it} + \eta_i + \nu_{it}
\end{equation}
with $\beta_{it}$ depending on $z_{it}$. We assume that each element of $\beta_{it}$, say $\beta_{itk}$, can be written as
$$\beta_{itk}=\tilde{z}'_{it}\gamma_k + a_{itk} \ ,$$
where $k \in \left\{1,2,3\right\}$, $a_{itk}$ a random term, and $\tilde{z}_{it}$ a spline basis evaluated at $z_{it}$. 
Given the data limitations it turned out that in our application it is sufficient if
$\tilde{z}_{it}$ consists of an intercept, $z_{it}$, and the squared values. When stacking the $\beta$-coefficients we obtain
\begin{eqnarray}\label{regression_beta}
\begin{pmatrix} \beta_{1it} \\ \beta_{2it} \\ \beta_{3it} \end{pmatrix}=
\begin{pmatrix}  \tilde{z}'_{it} & 0 & 0 \\ 0 & \tilde{z}'_{it} & 0 \\
0 & 0 &  \tilde{z}'_{it} \end{pmatrix}
\begin{pmatrix} \gamma_1 \\ \gamma_2 \\ \gamma_3 \end{pmatrix} +
\begin{pmatrix} a_{it1} \\ a_{it2} \\ a_{it3} \end{pmatrix} \in \mathbb{R}^{3} && , \\ \nonumber
\begin{pmatrix}  \tilde{z}'_{it} & 0 & 0 \\ 0 & \tilde{z}'_{it} & 0 \\
0 & 0 &  \tilde{z}'_{it} \end{pmatrix} =Z_{it} \in \mathbb{R}^{3 \times M} \ , \ 
\begin{pmatrix} \gamma_1 \\ \gamma_2 \\ \gamma_3 \end{pmatrix} = \gamma \in \mathbb{R}^M \ 
\mbox{and} \ \begin{pmatrix} a_{it1} \\ a_{it2} \\ a_{it3} \end{pmatrix} =a_{it} \in\mathbb{R}^3&&. 
\end{eqnarray}
We will assume that
\begin{eqnarray}\label{corr_a}
E(a_{it}a'_{js}) &=& \left\{\begin{array}{cl} \Lambda , \ &\mbox{if} \ i=j \ \mbox{and} \ s=t \ \mbox{and} \\ 0,   &\mbox{else.} \end{array}\right. 
\\ \nonumber
 \Lambda 
&= & \begin{pmatrix} Var(\beta_{it1})&Cov(\beta_{it1},\beta_{it2})&Cov(\beta_{it1},\beta_{it3})\\
Cov(\beta_{it2},\beta_{it1})&  Var(\beta_{it2}) &Cov(\beta_{it2},\beta_{it3})\\
Cov(\beta_{it3},\beta_{it1}) &Cov(\beta_{it3},\beta_{it2})&  Var(\beta_{it3}) \end{pmatrix}
\ \mbox{for all} \ i \ \mbox{and}  \  t. 
\end{eqnarray}
what just means that the correlations of the coefficients do not change over time and across individuals given $\zeta_{it}$.
Stacking the time-series data of \eqref{regression_vc} you can write
$$ y_i = \rho y_{i(-l)} + \eta_i \iota_{T-l} + X_i \beta_i + \nu_i \in \mathbb{R}^{T-l} \ , $$
with
\begin{eqnarray*} 
&& y_{i(-l)} = ( y_{i1}, \ldots ,  y_{i(T-l)})' \in \mathbb{R}^{T-l}, \
\iota_{T-l} =(1, \ldots, 1)' \in \mathbb{R}^{T-l}, \\
&& X_i = \begin{pmatrix} x_{i(1+l)}'    &    &    \\
  &       \ddots &   \\
  &       & x_{iT}' \end{pmatrix} \in \mathbb{R}^{(T-l) \times 3(T-l)},\\
&& \beta_i = ( \beta_{i(1+l)}', \ldots ,  \beta_{iT}')' \in \mathbb{R}^{3(T-l)} \ \mbox{and} 
\nu_i = ( \nu_{i(1+l)}, \ldots ,  \nu_{iT})' \in \mathbb{R}^{T-l} .
\end{eqnarray*}
Furthermore, we can stack the time-series data of \eqref{regression_beta} as
$$ \beta_i = Z_i \gamma + a_i \in \mathbb{R}^{3(T-l)} \ , $$
with
\begin{equation*} 
Z_i = (Z_{i(1+l)}', \ldots, Z_{iT}')' \in \mathbb{R}^{3(T-l) \times M} \ \mbox{and} \ 
a_i = (a_{i(1+l)}', \ldots , a_{iT}')' \in \mathbb{R}^{3(T-l)} . 
\end{equation*}
Furthermore, it follows from \eqref{corr_a}
$$ E(a_i a_i')=I_{T-l} \otimes \Lambda \in \mathbb{R}^{3(T-l) \times 3(T-l)} $$
for the random deviations from the conditional mean coefficients, and
$ E(\nu_i \nu_i')= \Sigma $ for further model deviations.
After stacking-time series data of \eqref{regression_vc}, 
we additionally stack cross-sectional data
\begin{equation}\label{regression_vc_stacked}
y=\rho y_{-l} + C\eta + X \beta + \nu ,
\end{equation}
with 
\begin{eqnarray*} 
&& y_{-l}=(y_{1(-l)}', \ldots , y_{n(-l)}')' \in \mathbb{R}^{n(T-l)}, \\
&& C =I_{T-l} \otimes \iota \in \mathbb{R}^{n(T-l) \times n} , \
\eta =(\eta_1, \ldots , \eta_n)' \in \mathbb{R}^n, \\
&& X = \begin{pmatrix} X_1   &    &    \\
   & \ddots   &    \\
    &    &    X_n \end{pmatrix} \in \mathbb{R}^{n(T-l) \times 3n(T-l)}, \\
&& \beta = (\beta_1', \ldots , \beta_n')' \in \mathbb{R}^{3n(T-l)} \ \mbox{and} \ 
\nu = (\nu_1', \ldots , \nu_n')' \in \mathbb{R}^{n(T-l)} .
\end{eqnarray*}
Furthermore, after stacking time-series data of \eqref{regression_beta}, we additionally stack cross-sectional data
\begin{equation}\label{regression_beta_stacked}
\beta= Z \gamma + a \in \mathbb{R}^{3n(T-l)},
\end{equation}
with 
\begin{equation*} 
Z =(Z_1', \ldots , Z_n')' \in \mathbb{R}^{3n(T-l) \times M} \ \mbox{and} \
a = (a_1', \ldots , a_n')' \in \mathbb{R}^{3n(T-l)}  .
\end{equation*}
When plugging \eqref{regression_beta_stacked} into \eqref{regression_vc_stacked} you get
\begin{eqnarray*}
y&=& \rho y_{-l} + C\eta + X \beta + \nu =\rho y_{-l} + C\eta + 
 \begin{pmatrix} X_1 &       &    \\
 &    \ddots   &    \\
 &       &    X_n \end{pmatrix}
\left[ \begin{pmatrix}  Z_1 \\ Z_2 \\ \vdots \\ Z_n    \end{pmatrix} \gamma + 
\begin{pmatrix}  a_1 \\ a_2 \\ \vdots \\ a_n    \end{pmatrix} \right]
+\nu \\
&=& \rho y_{-l} + C\eta + 
\begin{pmatrix}  X_1 Z_1 \\ X_2 Z_2 \\ \vdots \\ X_n Z_n    \end{pmatrix} \gamma +
\begin{pmatrix}  X_1 a_1 \\ X_2 a_2 \\ \vdots \\ X_n a_n    \end{pmatrix} + \nu \\
&=& \rho y_{-l} + C\eta + W \gamma + u \in \mathbb{R}^{n(T-1)} ,
\end{eqnarray*}
where we used the notation 
\begin{equation*}
W =(Z_1'X_1', \ldots, Z_n' X_n')' \in \mathbb{R}^{n(T-l) \times M} \ \mbox{and} \
u =(a_1'X_1',a_2'X_2', \ldots,a_n'X_n')' + \nu \in \mathbb{R}^{n(T-l)} .
\end{equation*}
The regression equation
\begin{equation} \label{reg_interest}
y=\rho y_{-l} + C\eta + W \gamma + u \in \mathbb{R}^{n(T-l)}
\end{equation}
has $n+M+1$ parameters. If the matrix
$$ (y_{-l},C,W) \in \mathbb{R}^{n(T-l) \times (n+M+1)} $$
has full column rank, the model is identified. The following calculation shows how the
unobserved heterogeneity causes uncorrelated but heteroscedastic errors if the $\nu_{it} $
are uncorrelated:
\begin{eqnarray*}
E\left[uu'\right]&=& E \left[\left( \begin{pmatrix} X_1 a_1 \\ X_2 a_2 \\ \vdots \\ X_n a_n \end{pmatrix} + \nu  \right)
\left( \begin{pmatrix} X_1 a_1 \\ X_2 a_2 \\ \vdots \\ X_n a_n \end{pmatrix} + \nu \right)' \right] \in \mathbb{R}^{n(T-l) \times n(T-l)}   \\
&=& \begin{pmatrix} X_1E(a_1a_1')X_1' &  &   &  \\
  & X_2E(a_2a_2')X_2' &   &    \\
  &  & \ddots  &    \\
  &  &  & X_nE(a_na_n')X_n'  \end{pmatrix} + \Sigma   \\
&=&  \begin{pmatrix} x_{1(1+l)}'\Lambda x_{1(1+l)} & & & & & & \\
    & \ddots & & & & & \\   
    & & x_{1T}'\Lambda x_{1T}& & & & \\
    & & & \ddots & & & \\ 
    & & & & x_{n(1+l)}'\Lambda x_{n(1+l)} & & \\ 
    & & & & & \ddots & \\  
    & & & & & & x_{nT}'\Lambda x_{nT}   \end{pmatrix} + \Sigma \  . 
\end{eqnarray*}
Obviously, under the standard assumptions for random coefficient models, OLS gives still a consistent but inefficient estimator.
In a similar context, \citet{amemiya_1978} proposes to start with estimating the $\beta$'s from
$$ y=\rho y_{(-l)} + C\eta +
\begin{pmatrix} X_1 \\ 0 \\ \vdots \\ 0 \end{pmatrix} \beta_1 +
\begin{pmatrix} 0 \\ X_2 \\ \vdots \\ 0 \end{pmatrix} \beta_2 +
\ldots +
\begin{pmatrix} 0 \\ 0 \\ \vdots \\ X_n \end{pmatrix} \beta_n +
\nu \in \mathbb{R}^{n(T-1)}
$$
and afterward the $\gamma$, $\Lambda$ and $\Sigma$ using the $\beta$ estimates. 
Even with $\Sigma = \sigma^2 I_{n(T-l)}$ his would mean to estimate $3n(T-l) + n +1$ parameters with only $n(T-l)$ data what can't work. 
A feasible way to obtain an estimator is to make use of the linear structure of $E(uu')$ and formulate an auxiliary regression in the following way. You first apply OLS to \eqref{reg_interest} to obtain consistent estimators for $\rho$, $\eta$ and $\gamma$. Then, extract the residuals from the regression and regress the squared residuals on the variables given in the linear structure of $E(uu')$ to estimate $\sigma^2$ (if $\nu$ homoscedastic; extensions are obvious), and $\Lambda$. This can be done using the quadratic programming following \citet{goldfarb_1982,goldfarb_1983}. 
The reciprocal fitted values of this regression can be used as weights to estimate the coefficients of \eqref{reg_interest}. Then you iterate until the estimated coefficients $\rho$, $\eta$ and $\gamma$ converge. This method has the problem that, when having estimated the residuals of \eqref{reg_interest} in one step, one has to find the $\lambda$'s such that the matrix $\Lambda$ has the characteristics of a covariance matrix (symmetric and positive definite). Applying OLS on the auxiliary regression does not guarantee this and may result in negative weights. Estimating the Cholesky decomposition of the matrix $\Lambda$ is not possible because of the resulting multicollinearity. Incorporating the symmetry condition is easy, and it is also possible to formulate inequality conditions for the $\lambda$'s such that $\Lambda$ fulfills some of the characteristics of a covariance matrix, e.g.\ to force the diagonal elements of $\Lambda$ to be positive. However, the resulting optimization procedure that calculates the $\lambda$'s itself 
has errors and one cannot be sure that the result of iterated least-squares combined with the iterated solution of the optimization procedure in every step converges to the desired result. 
Therefore, we propose to estimate the coefficients of \eqref{reg_interest} as follows: 
\begin{itemize}
\item[(1)] We estimate the coefficients of \eqref{reg_interest} in the first step using OLS, 
\item[(2)] we extract the residuals, 
\item[(3)] we estimate the coefficients of \eqref{reg_interest} again using least-squares with the reciprocal squared residuals as weights.
\end{itemize}
Extracting the residuals from this regression (repeat step (2)) gives weights for the next regression (repeat step (3)) and so on. We iterate this procedure, until the sum of squared differences of the coefficients from one step to the next is smaller than $0.005$. This ensures that the average squared difference from one step to another is approximately $0.00005$. 

\section{Results}
\label{sec:results}

\subsection{The Effects on Economic Growth}
\label{subsection:results_growth}

In this subsection, we investigate the effects of poverty, inequality and the middle class on the coefficients of our growth equation. We will consider three year economic growth. Considering too short term economic growth might lead to a spurious regression problem and endogeneity, while we lose observations by considering too long term economic growth. We found three years to be reasonable, but we admit that this choice is arbitrary. More specifically, we estimate a lagged regression equation with $l=3$ and $x_{it}=(lnn_{i(t-3)},lnsk_{i(t-3)},lnattain_{i(t-3)}) \in \mathbb{R}^3$. Furthermore, $$ \tilde{z}_{it}=(1,pov_{i(t-3)},pov_{i(t-3)}^2, gini_{i(t-3)},gini_{i(t-3)}^2, middleclass_{i(t-3)}, middleclass_{i(t-3)}^2 )' \in \mathbb{R}^7 , $$ which implies $M=21$. In this case the time-series covers $31$ years, namely from the year 1973 to the year 2003.
 
\begin{figure}[!ht]
\centering
\subfloat{\includegraphics[width=0.49\textwidth,keepaspectratio]{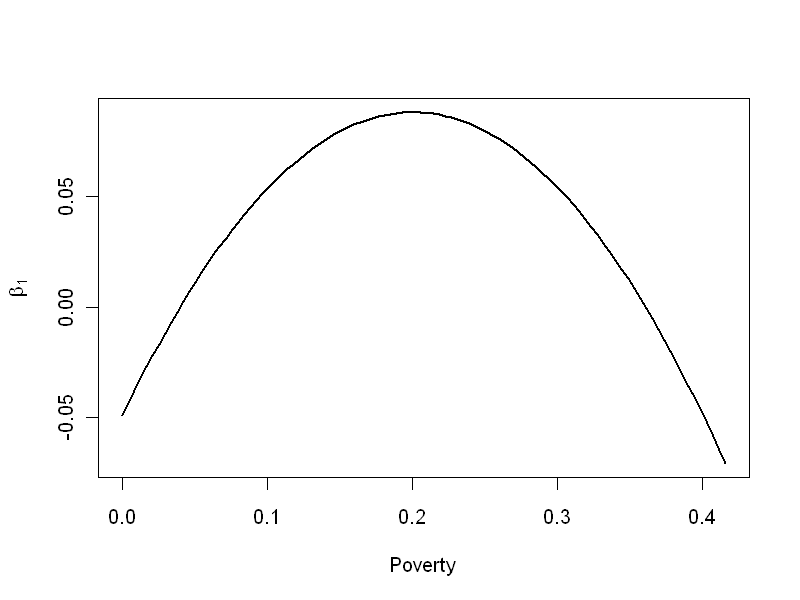}}
\subfloat{\includegraphics[width=0.49\textwidth,keepaspectratio]{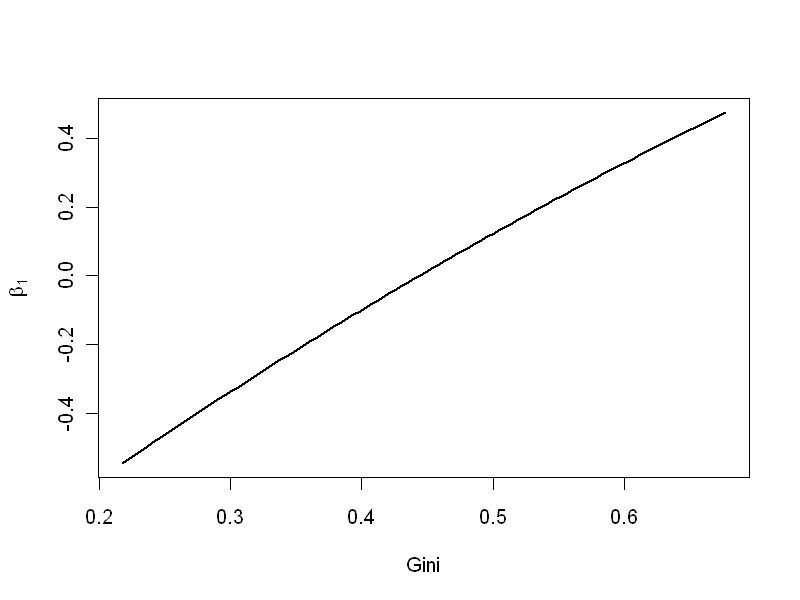}} \\
\subfloat{\includegraphics[width=0.49\textwidth,keepaspectratio]{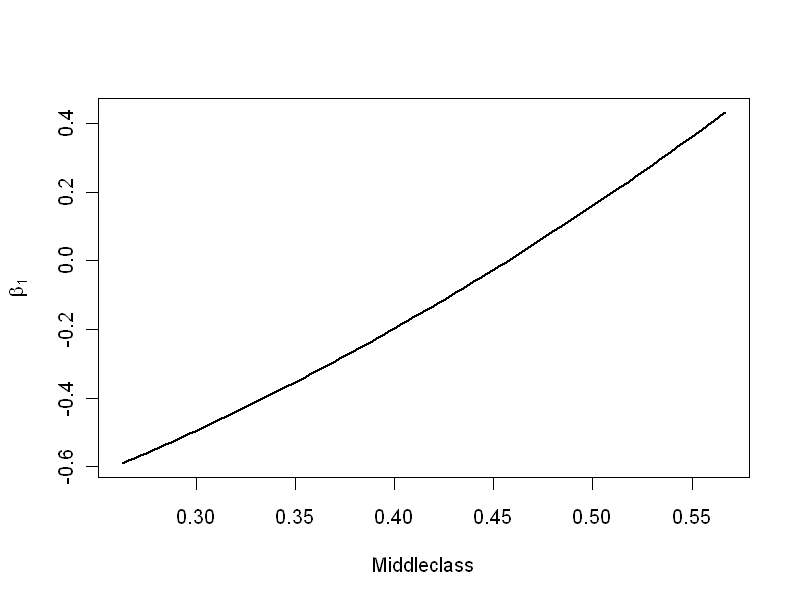}}
\subfloat{\includegraphics[width=0.49\textwidth,keepaspectratio]{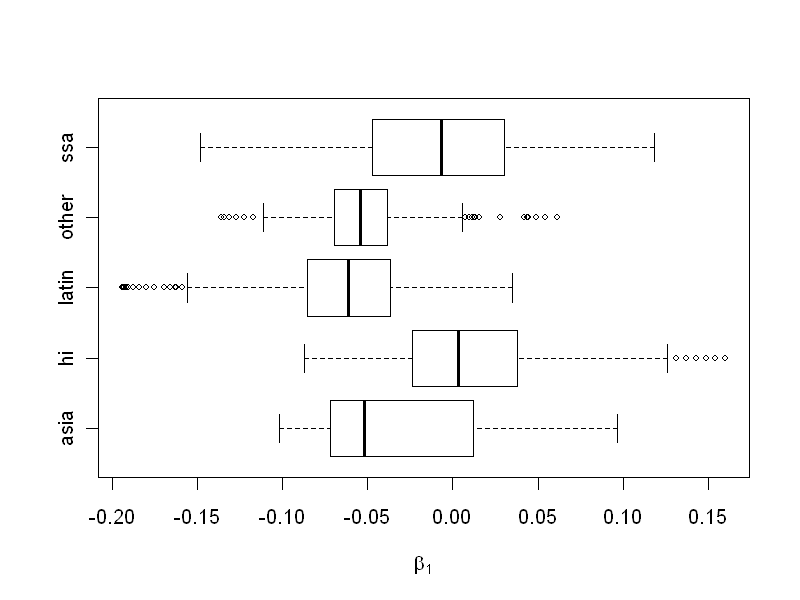}}
\caption{The effects of poverty, inequality and the middle class on $\beta_{1}$ and the $\beta_{1}$'s stratified for the groups of countries. The maximum widths of the confidence bands are $0.0118$ for poverty, $0.02306$ for gini and $0.02432$ for middle class.}
\label{figure:beta_lnn}
\end{figure}

When displaying the estimated coefficients, we report the level of significance of the coefficient ${***}$ if the p-value is almost zero, ${**}$ if the p-value is smaller than $0.01$, and ${*}$ if the p-value is smaller than $0.05$. The estimated autoregressive coefficient is $0.9318^{***}$.\\ 
The regression that explains $\beta_1$, which is the coefficient of $lnn$ is
\begin{equation*}
\begin{split}
\beta_{1it}=&-2.1952^{***}+ 1.3705^{***} pov_{i(t-3)} -3.4198^{***} pov_{i(t-3)}^2 + 2.9553 gini_{i(t-3)}^{***}\\
&-0.8132^{***} gini_{i(t-3)}^2 
+0.8917^{***} middleclass_{i(t-3)} + 2.9860^{***} middleclass_{i(t-3)}^2.
\end{split}
\end{equation*}
A graphical illustration of this is given in figure \eqref{figure:beta_lnn}. The plots show the evolution of $\beta_1$ along $pov$ (upper left), $gini$ (upper right) and $middleclass$ (bottom left). For the variables that are hold constant in each plot, we plug in its observed averaged. Differences in inequality and the income earned by the middle class have a much larger impact on $\beta_1$ than differences in the poverty rate. The relationship of $gini$ and $middleclass$ to $\beta_1$ is similar, namely almost linear and increasing. The poverty rate has an inverted U-shaped relationship. The returns to $lnn$ are theoretically the largest in countries, where inequality is large even though the middle class earns a high fraction of the total income and a serious fraction of the total population (approximately 20\%) earns below the poverty line. The boxplots (bottom right) show the estimated coefficients stratified for different country groups. The groups are Asia, Latin, sub-Saharan Africa (SSA), High Income (HI, the High Income OECD and the High Income Non-OECD) and the group of other countries (the Middle East and North Africa and the Eastern Europe). We modified the World Bank grouping appropriately for our data. The returns to $lnn$ are especially large for sub-Saharan African and HI countries. For the sub-Saharan Africa group, this result coincides with \citet{koehler_2011}. We also show that the $\beta_1$-coefficients of sub-Saharan African countries have larger variation than other countries. The coefficients of the groups Latin and Other are smaller on average and have less variation. It is interesting to see that not only for HI and sub-Saharan African but also for other countries the returns to $lnn$ can be positive. The overall distribution of the variable-coefficients seems to be in accordance with the mean-coefficients model, by which the return to $lnn$ is estimated to be $-0.00640$.

\begin{figure}[!ht]
\centering
\subfloat{\includegraphics[width=0.49\textwidth,keepaspectratio]{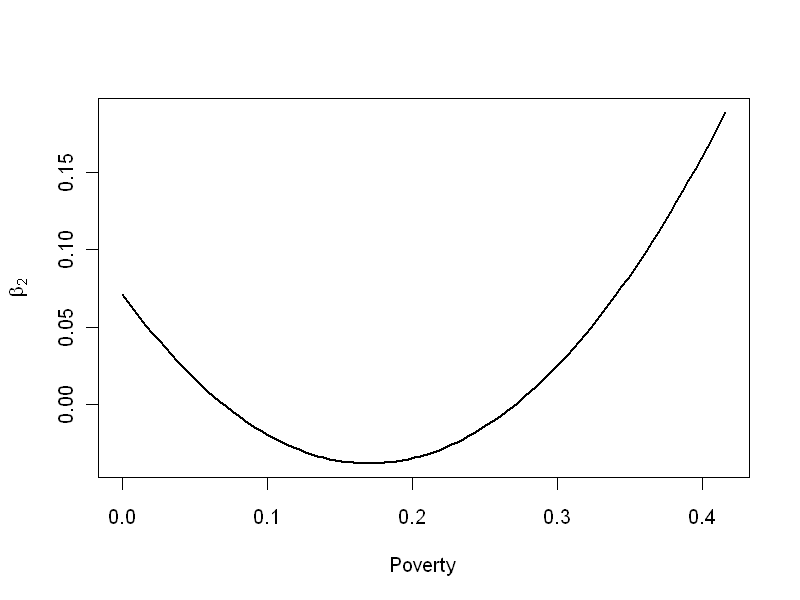}}\subfloat{\includegraphics[width=0.49\textwidth,keepaspectratio]{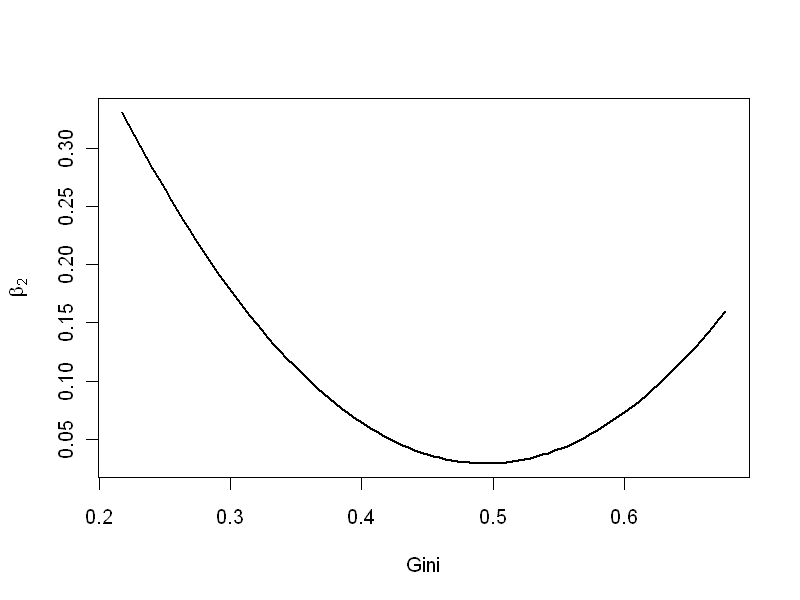}}\\
\subfloat{\includegraphics[width=0.49\textwidth,keepaspectratio]{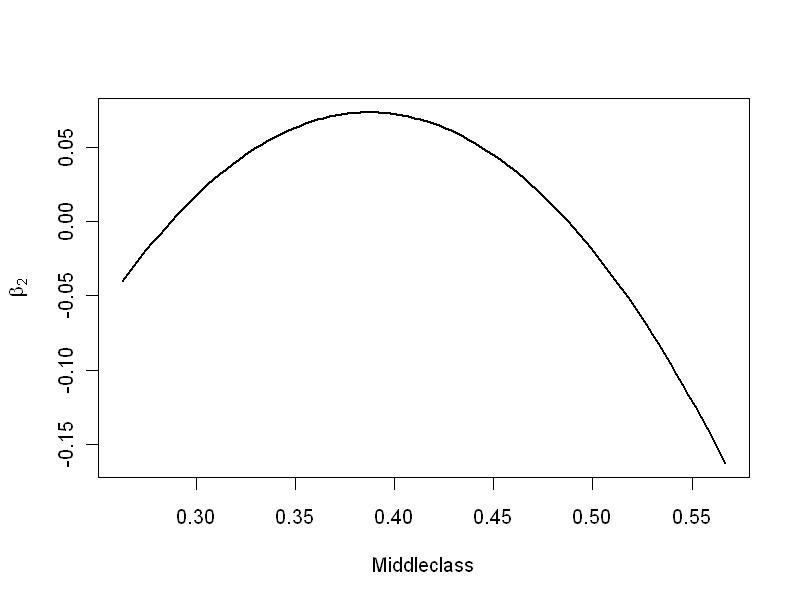}}
\subfloat{\includegraphics[width=0.49\textwidth,keepaspectratio]{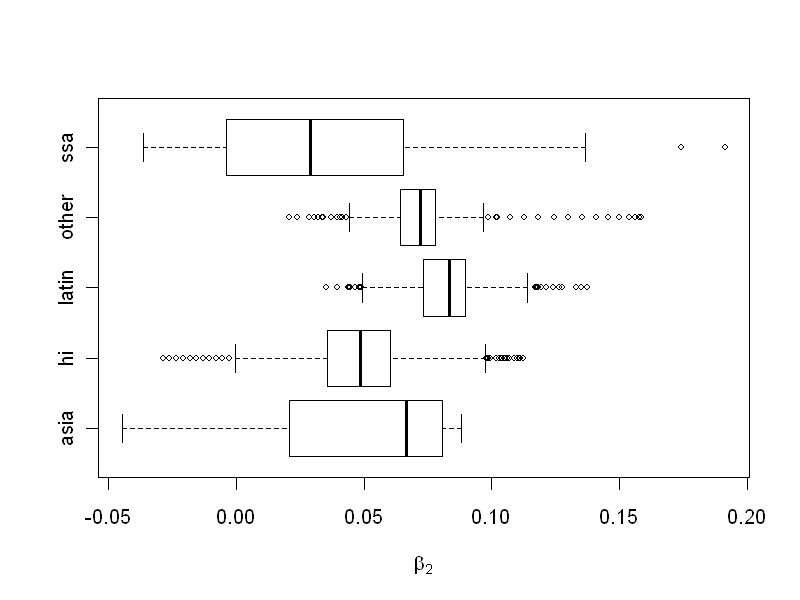}}
\caption{The effects of poverty, inequality and the middle class on $\beta_{2}$ and the $\beta_{2}$'s stratified for the groups of countries. The maximum widths of the confidence bands are $0.02387$ for poverty, $0.01866$ for gini and $0.02138$ for middle class.}
\label{figure:beta_lnsk} 
\end{figure}

The estimated regression equation that explains the coefficient of $lnsk$ is
\begin{equation*}
\begin{split}
\beta_{2it}=&-0.0519^{**} -1.2799^{***} pov_{i(t-3)} + 3.7563^{***} pov_{i(t-3)}^2 -3.8873^{***}  gini_{i(t-3)} \\
&+ 3.9307^{***} gini_{i(t-3)}^2 
+5.6845^{***} middleclass_{i(t-3)} -7.3372^{***} middleclass_{i(t-3)}^2,
\end{split}
\end{equation*}
which is plotted in figure \eqref{figure:beta_lnsk}. It can be observed that poverty, inequality and the middle class have a remarkable impact on the return to investment in physical capital. We observe a U-shaped relationship for the variables $pov$ and $gini$, and an inverted U-shaped relationship for $middleclass$. Therefore, the returns to investments are the highest, when poverty and inequality are either quite low or large while the middle class earns about 40\% of the total income. The boxplots of figure \eqref{figure:beta_lnsk} show that sub-Saharan Africa has the smallest returns to physical capital on average. This is here explained by large inequality, small middle class and large poverty. The coefficients of sub-Saharan African countries are also subject to large variation. The coefficients of Asia are small on average and show that the underlying distribution is skewed. The group Other has smaller variation and larger coefficients on average. The largest returns to physical capital are observed for Latin American countries. These countries are characterized by small poverty rates but extreme inequality und a moderate size of middle class. When considering the mean-coefficients model, the return to $lnsk$ is estimated as $0.04919$, which seems to be coherent with the our estimates.

\begin{figure}[!ht]
\centering
\subfloat{\includegraphics[width=0.49\textwidth,keepaspectratio]{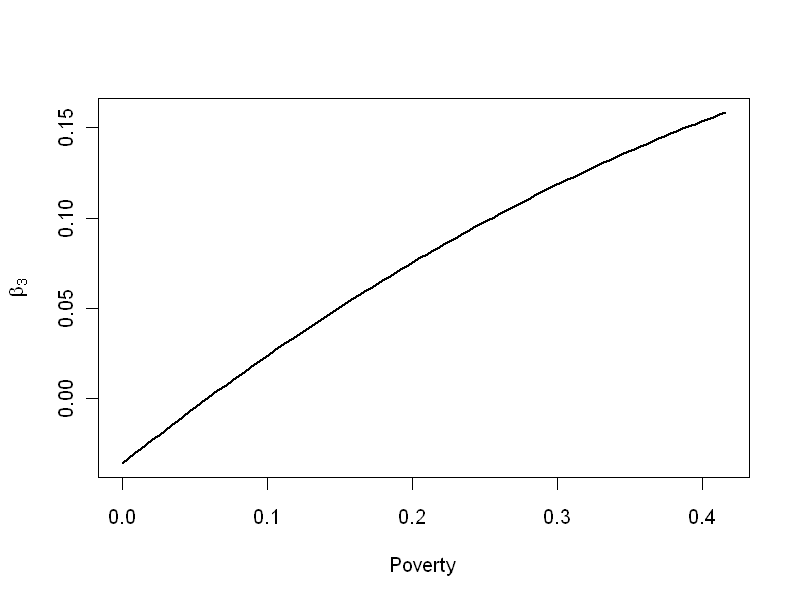}}\subfloat{\includegraphics[width=0.49\textwidth,keepaspectratio]{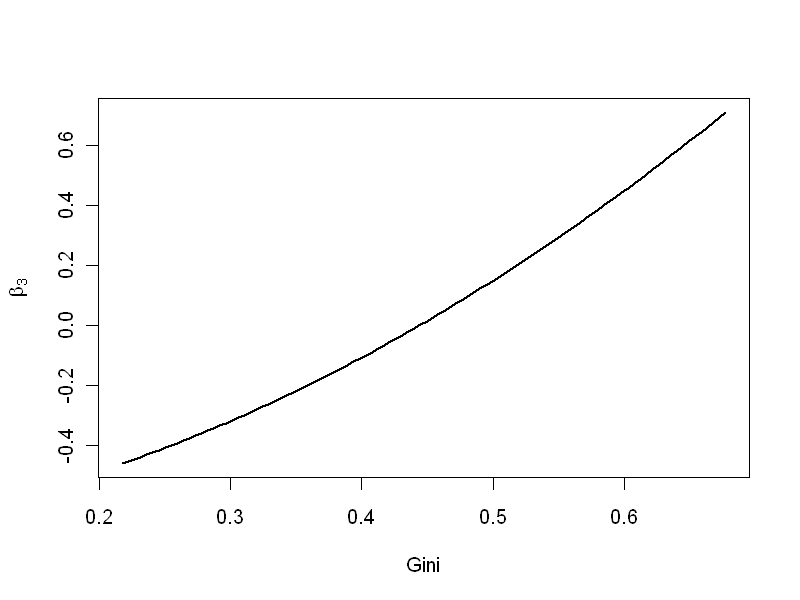}}\\
\subfloat{\includegraphics[width=0.49\textwidth,keepaspectratio]{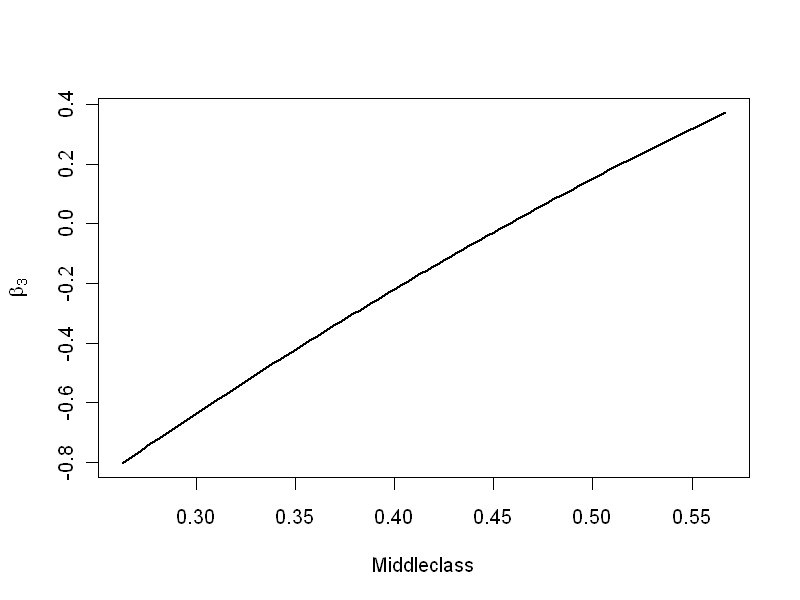}}
\subfloat{\includegraphics[width=0.49\textwidth,keepaspectratio]{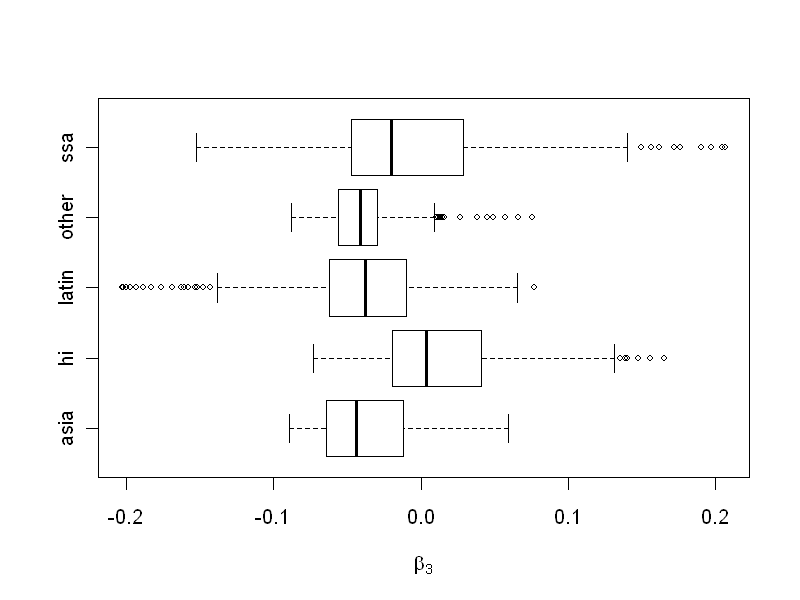}}
\caption{The effects of poverty, inequality and the middle class on $\beta_{3}$ and the $\beta_{3}$'s stratified for the groups of countries. The maximum widths of the confidence bands are $0.01695$ for poverty, $0.01745$ for gini and $0.01906$ for middle class.}
\label{figure:beta_lnattain} 
\end{figure}

The estimated regression equation for the coefficient of $lnattain$ is given by
\begin{equation*}
\begin{split}
\beta_{3it}=&-2.8512^{***} + 0.6369^{***} pov_{i(t-3)} -0.4078^{***} pov_{i(t-3)}^2 +0.5740^{***}  gini_{i(t-3)} \\
&+ 2.2057^{***} gini_{i(t-3)}^2 
+5.7917^{***} middleclass_{i(t-3)} -2.3057^{***} middleclass_{i(t-3)}^2.
\end{split}
\end{equation*}
The boxplots in figure \eqref{figure:beta_lnattain} shows that the majority of the estimated coefficients of $lnattain$ are negative. The mean-coefficients model estimates the return to $lnattain$ to be $-0.03239$, which is again negative. This counterintuitive result might be explained by a measurement problem or related to the functional misspecification of the simple linear growth model. Alternatively, \citet{pritchett_1996} provides theories why schooling may not lead to economic growth. For example, the returns to education fall rapidly when the demand for educated labor is stagnant. We observe that $pov$, $gini$ and $middleclass$ have an almost linear and increasing relationship with $\beta_3$. High inequality, high poverty and a large share of income to the middle class causes a high return to school attainment. Differences in poverty have a much smaller impact on the coefficient than inequality or the middle class. 

The results clearly show that coefficients can differ a lot depending on poverty, inequality and the fraction of income earned by the middle class. By considering varying coefficients, we can also see the different patterns across country groups. Certainly, we expect less bias in the coefficient estimates since we have explicitely modeled different returns of the determinants of growth.

\subsection{The Effects on the Economic Growth of the Poor and the Rich} \label{subsection:results_growth_up_low}

We finally investigate the growth path for the upper and the lower twenty per cent of the society in each country. Differences in the growth path of the poor and the rich naturally affect the income distribution which in turn affect growth. Therefore, this exercise will provide us insights how the income distributions evolve. We use an analogous econometric procedure as in the previous subsection, except that the dependent variables are $y_{rich,it}$ and $y_{poor,it}$ defined in Section \ref{sec2}. We note that these dependent variables are based on GDP's per capita, having thus a different scaling for the dependent variable as in the previous subsection which was based on GDP per worker. We will avoid any comparison of the results based on different scalings. Instead, we focus on investigating how the income distribution measures affect the pro-poor and the pro-rich growth respectively.

\begin{figure}[!ht]
\centering
\subfloat{\includegraphics[width=0.49\textwidth,keepaspectratio]{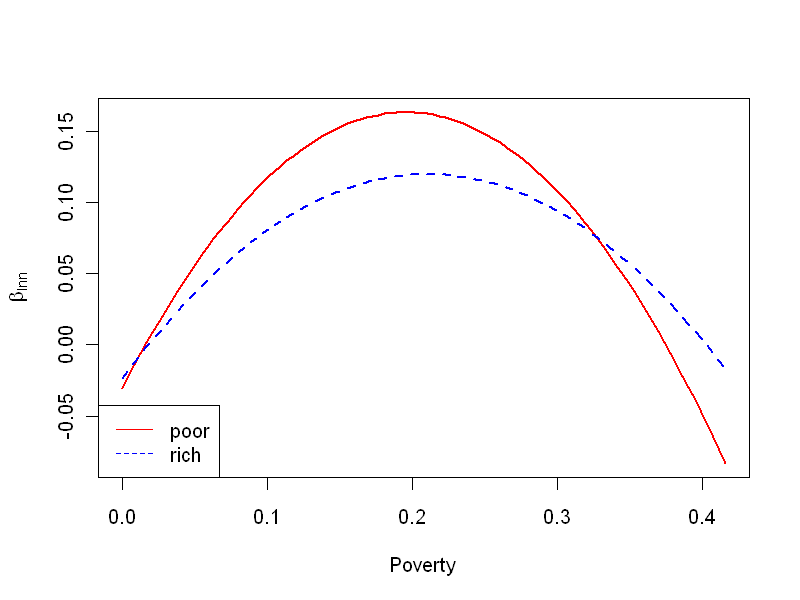}}\subfloat{\includegraphics[width=0.49\textwidth,keepaspectratio]{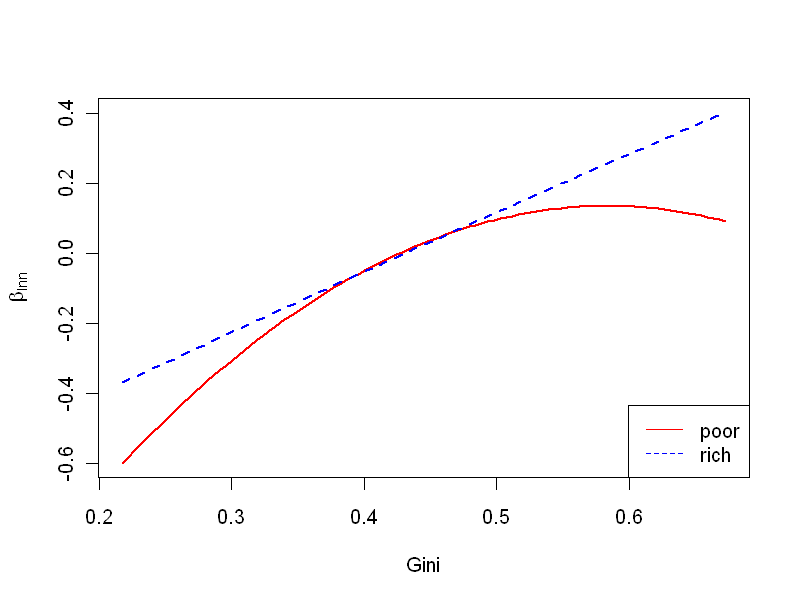}}\\
\subfloat{\includegraphics[width=0.49\textwidth,keepaspectratio]{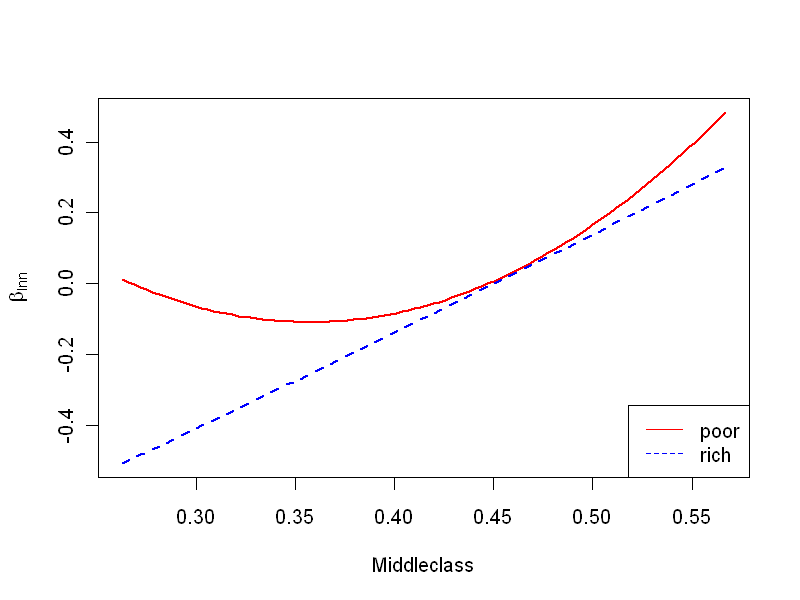}}
\caption{The effects of poverty, inequality and the middle class on $\beta_{1}$ of the poorest and richest twenty percent}
\label{figure:beta_lnn_pr} 
\end{figure}

The autoregressive coefficient for the poor is $0.9704^{***}$ and that for the rich is $0.9366^{***}$. This demonstrates that the series of the dependent variable is more persistent for the poor than for the rich.
The evolution of the return to $lnn$ of the poorest 20 percent is 
\begin{equation*}
\begin{split}
\beta_{1it, \ poor}=&-0.1806^{***} + 1.9801^{***} pov_{i(t-3)} -5.0660^{***} pov_{i(t-3)}^2  +6.4187^{***}  gini_{i(t-3)} \\
&-5.5026^{***} gini_{i(t-3)}^2
-9.6406^{***} middleclass_{i(t-3)} +13.4882^{***} middleclass_{i(t-3)}^2
\end{split}
\end{equation*}
and that of the richest twenty percent is given by
\begin{equation*}
\begin{split}
\beta_{1it, \ rich}=&-1.9793^{***} + 1.3614^{***} pov_{i(t-3)}  -3.2350^{***} pov_{i(t-3)}^2 +1.8338^{***}  gini_{i(t-3)}\\ 
&-0.1519^{***} gini_{i(t-3)}^2 
2.4768^{***} middleclass_{i(t-3)} +0.3237^{***} middleclass_{i(t-3)}^2.
\end{split}
\end{equation*}
This is graphically demonstrated in figure \eqref{figure:beta_lnn_pr}. The main drivers of the return to $lnn$ are inequality and the share of income of the middle class, while the poverty rate has hardly any influence on $\beta_1$. Poverty has an inverted U-shaped relationship with $\beta_1$ for the poor and the rich. Inequality and the share earned by the middle class have an almost linear and increasing relationship with $\beta_1$ for the rich. Inequality and the share of income of the middle class show an inverted U-shaped and an U-shaped relationship with $\beta_1$ respectively for the poor. The depreciation rate has large returns to economic growth of the rich with around 20\% population living under the poverty line, large inequality and large income share of the middle class. It is similar for the pro-poor growth, except that very high inequality ($gini$ above 0.6) suppresses the return, and very low income share of the middle class ($middleclass$ below 0.35) increases it.

\begin{figure}[!ht]
\centering
\subfloat{\includegraphics[width=0.49\textwidth,keepaspectratio]{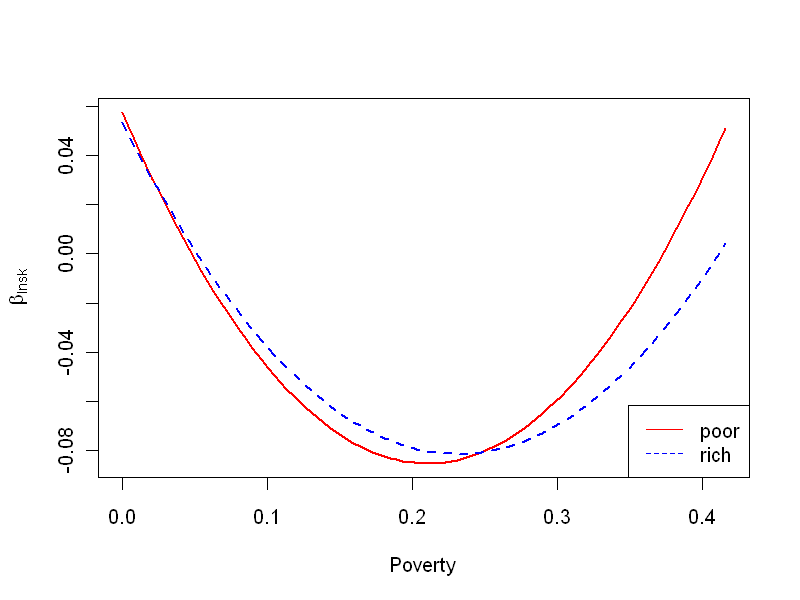}}\subfloat{\includegraphics[width=0.49\textwidth,keepaspectratio]{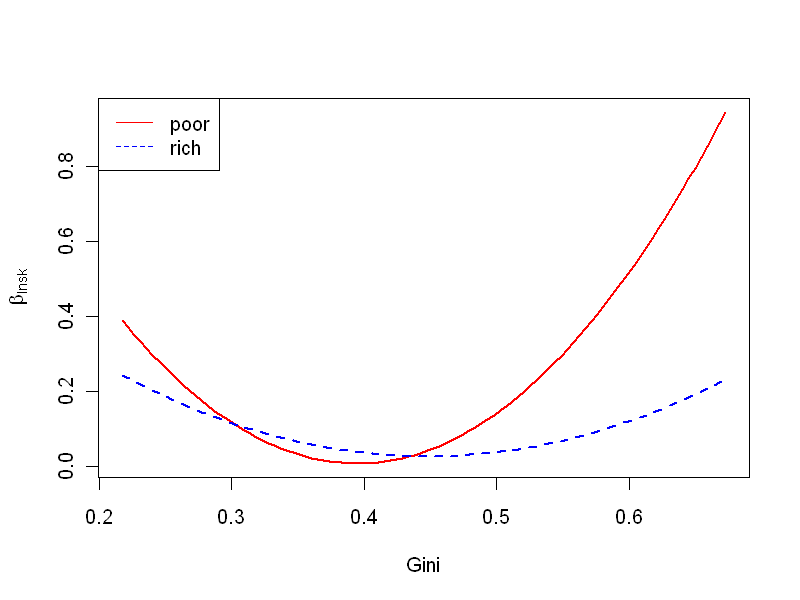}}\\
\subfloat{\includegraphics[width=0.49\textwidth,keepaspectratio]{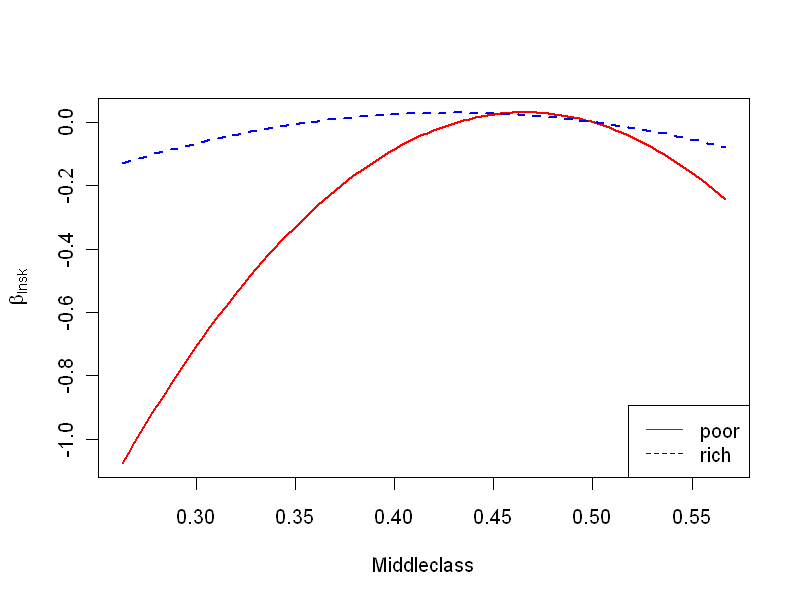}}
\caption{The effects of poverty, inequality and the middle class on $\beta_{2}$ of the poorest and the richest twenty percent}
\label{figure:beta_lnsk_pr} 
\end{figure}

The evolution of the return to $lnsk$ for the poorest 20 percent of the population is 
\begin{equation*}
\begin{split}
\beta_{2it}=&-3.9012^{***} -1.3569^{***} pov_{i(t-3)} +3.2247^{***} pov_{i(t-3)}^2  -9.5966^{***}  gini_{i(t-3)} \\
&+ 12.1551^{***} gini_{i(t-3)}^2 
+25.0670^{***} middleclass_{i(t-3)} -26.9079^{***} middleclass_{i(t-3)}^2
\end{split}
\end{equation*}
and for the richest twenty percent is
\begin{equation*}
\begin{split}
\beta_{2it}=&-0.1896^{***} -1.1640^{***} pov_{i(t-3)}  +2.5135^{***} pov_{i(t-3)}^2 -3.6437^{***}gini_{i(t-3)} \\
&+ 4.0661^{***} gini_{i(t-3)}^2 
+4.9471^{***} middleclass_{i(t-3)} -5.7624^{***} middleclass_{i(t-3)}^2.
\end{split}
\end{equation*}
This is graphically demonstrated in figure \eqref{figure:beta_lnsk_pr}. Poverty has smaller influence on $\beta_2$ than inequality and the income share of the middle class. The relationship between $\beta_{2}$ and poverty and between $\beta_{2}$ and inequality are U-shaped, while the relationship between $\beta_{2}$ and the income share of the middle class is inverted U-shaped. 
For both growth paths, the return to investment in physical capital is high when poverty and inequality are either very low or very high, and when the income share of the middle class is around 45\%. It is interesting that when inequality is very high ($gini$ around 0.65), the return to investment in physical capital is saliently higher for the poor compared to the rich. On the other hand, when the income share of the middle class is extremely low ($middleclass$ around 0.25), the return for the poor is saliently lower compared to the rich. 

\begin{figure}[!ht]
\centering
\subfloat{\includegraphics[width=0.49\textwidth,keepaspectratio]{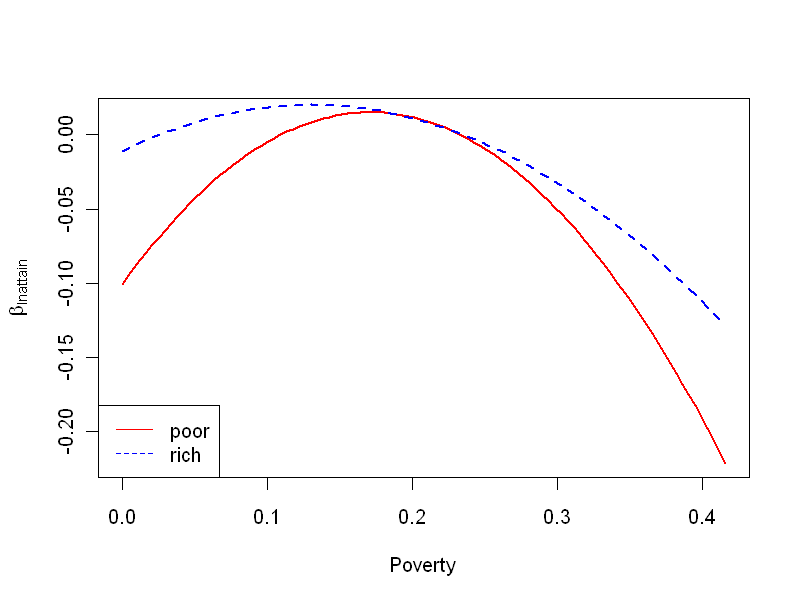}}\subfloat{\includegraphics[width=0.49\textwidth,keepaspectratio]{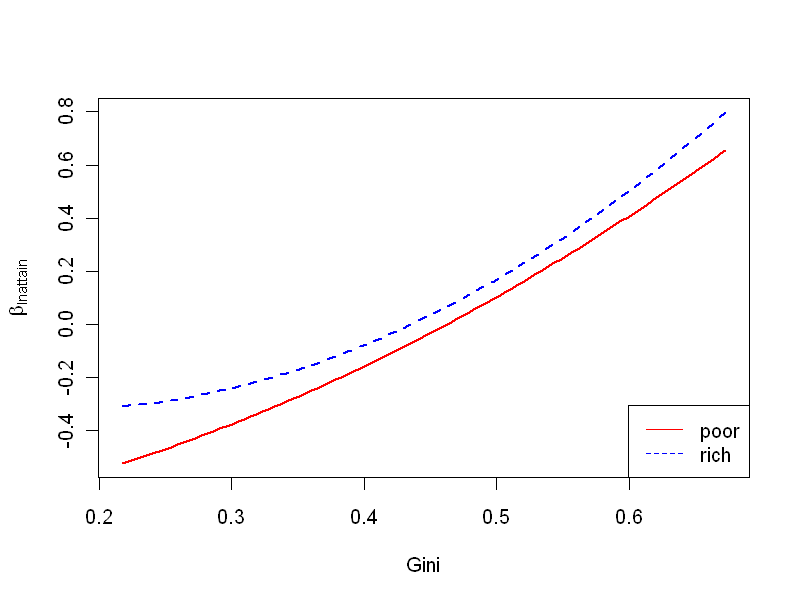}}\\
\subfloat{\includegraphics[width=0.49\textwidth,keepaspectratio]{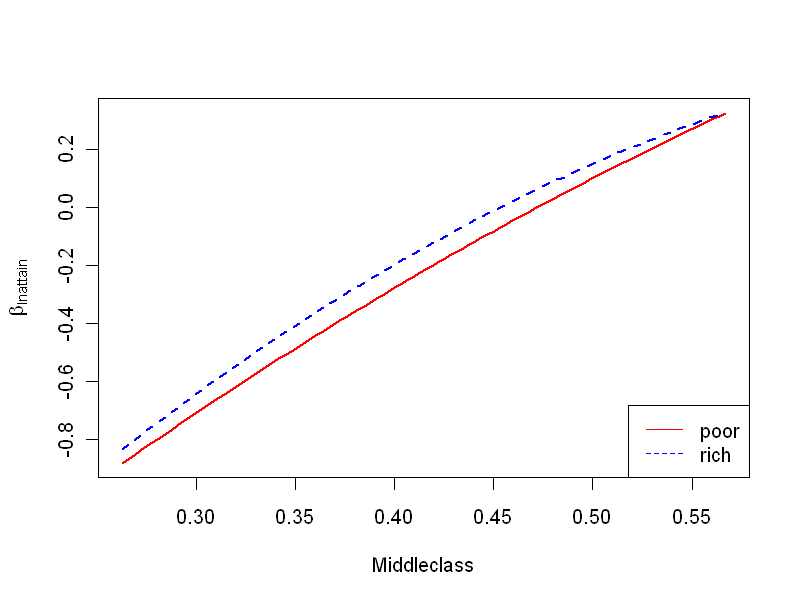}}
\caption{The effects of poverty, inequality and the middle class on $\beta_{3}$ of the poorest and the richest twenty percent}
\label{figure:beta_lnattain_pr} 
\end{figure}

The evolution of the coefficient of $lnattain$ for the poorest 20 percent of the population is
\begin{equation*}
\begin{split}
\beta_{3it}=&-3.0506^{***} +1.3524^{***} pov_{i(t-3)} -3.9498^{***} pov_{i(t-3)}^2  +0.6753^{***}  gini_{i(t-3)}\\
&- 2.1525^{***} gini_{i(t-3)}^2 
6.1614^{***} middleclass_{i(t-3)} -2.6327^{***} middleclass_{i(t-3)}^2
\end{split}
\end{equation*}
and for the richest twenty percent is given by
\begin{equation*}
\begin{split}
\beta_{3it}=&-2.8011^{***} +0.4760^{***} pov_{i(t-3)} -1.8265^{***} pov_{i(t-3)}^2 -1.3797^{***}  gini_{i(t-3)} \\
&+ 4.2849^{***} gini_{i(t-3)}^2
+ 7.9466^{***} middleclass_{i(t-3)} -4.9629^{***} middleclass_{i(t-3)}^2.
\end{split}
\end{equation*}
This is graphically demonstrated in figure \eqref{figure:beta_lnattain_pr}. Poverty has smaller influence on $\beta_{3}$ than inequality and the income share of the middle class. Analogous to subsection \eqref{subsection:results_growth}, we observe that the coefficients are likely to be negative. The relationship between $\beta_{3}$ and poverty is inverted U-shaped, while they are almost linear and increasing for inequality and the income share of the middle class. The return to investment in human capital is large with high inequality and high income share of the middle class and the poverty rate being around 20\% (for the poor and rich alike). Investments in human capital are slightly better for the rich in case of an extremely small or extremely large poverty rates.

The aforementioned results show that economic growth for the poor and the rich are differently influenced by the income distribution. These different economic growth paths of the poor and the rich have again implications to the income distribution.

\section{Conclusion}\label{sec:conlcusion}

Collecting long time-series for a large range of countries allows applying a varying coefficient model in which differences to the mean coefficient are explained by poverty, inequality and the share of income earned by the middle class. The results show that the returns exhibit a serious 
heterogeneity with respect to these factors. It is clear that neglecting this by estimating
a simple linear model leads to inconsistent estimators; but more importantly, the researcher
misses some of the maybe most important variation.  

There are several reasons for considering varying coefficients. First, adding measures of the income distribution to the set of explanatory variables of the growth regression alone, does not model 
their impact, and it ignores the fact that the poor behave different than the rich. Furthermore, one might lose a sensible economic interpretation when simply adding additively a lot of variables to the classic growth regression. Often, the mean of the coefficients is not an informative parameter of the growth equation because of the dramatic heterogeneity of returns to production factors. Furthermore, the differences of the coefficients to their means are highly suspicious to move simultaneously with growth, which indicates an endogeneity problem. Finally, as poor countries have weaker databases and are therefore more likely to be excluded from the data, the mean coefficient might suffer from a sample selection bias when simply estimated from fixed coefficient models. 

Remarkable results are that sub-Saharan African countries have highly varying and large returns to population growth, and highly varying but small returns to physical capital. Latin American countries experience highly negative returns to population growth but large positive returns to physical capital. All country groups experience negative returns to school attainment, which indicates that the variable does not take important information, such as quality of schooling into account. When expressing the coefficients as functions of poverty, inequality and the share of income earned by the middle class, then we observe that poverty has much smaller effects on the returns than the other two factors. Large inequality usually goes hand in hand with a small share earned by the middle class and vice versa. The fact that this tends to move the coefficients in opposite directions demonstrates the importance of incorporating both variables.

We also investigate the growth path of the poorest and the richest twenty percent of the society
respectively. Again we observe that the returns of the growth regression are highly dependent on poverty, inequality and the share earned by the middle class. Interestingly, the returns of the two subgroups of the total population are impacted in different ways. Outstanding results are that in case of extremely high and extremely small inequality, the return to population growth is smaller for the poor than for the rich, whereas in case of an extremely small share earned by the middle class (or an extremely large share earned by the middle class), it is larger. Furthermore, in case of extremely large inequality the return to investment in physical capital is larger for the poor, but in case of an extremely small share earned by the middle class it is smaller. This shows again that small inequality and a large share earned by the middle class (or large inequality and a small share earned by the middle class) tend to force the returns in different directions. The aforementioned differences of the poor and the rich naturally affect the parameters of the income distribution, which in turn affect the growth path of the GDP per worker. This undermines again the importance of considering the income distribution when modeling the growth path.


\begin{thebibliography}{}

\bibitem[\protect\citeauthoryear{Acemoglu and Zilibotti}{1997}]{acemoglu_1997} \textsc{Acemoglu, D. and Zilibotti, F.} (1997). Was Prometheus Unbound by Chance? Risk, Diversification, and Growth.
\textit{Journal of Political Economy} \textbf{105}(4) 709--751.

\bibitem[\protect\citeauthoryear{Adelman and Morris}{1969}]{adelman_1969} \textsc{Adelman, I. and Morris, C. T.} (1969). Society, Politics, and Economic Development: A Quantitative Approach.
\textit{The Economic Journal} \textbf{79}(313) 160--163.

\bibitem[\protect\citeauthoryear{Aghion and Bolton}{1997}]{aghion_1997} \textsc{Aghion, P. and Bolton, P.} (1997). A Theory of Trickle-Down Growth and Development.
\textit{Review of Economic Studies} \textbf{64}(2) 151--172.

\bibitem[\protect\citeauthoryear{Alesina and Rodrik}{1994}]{alesina_1994a} \textsc{Alesina, A. and Rodrik, D.} (1994). Distributive Politics and Economic Growth.
\textit{Quarterly Journal of Economics} \textbf{109}(2) 465--490.

\bibitem[\protect\citeauthoryear{Alesina}{1994}]{alesina_1994b} \textsc{Alesina, A.} (1994). Voting for Reform: Democracy, Political Liberalization, and Economic Adjustment: Democracy, Liberalization and Economic Adjustment (Chapter 2).
\textit{Oxford University Press}

\bibitem[\protect\citeauthoryear{Alesina and Perotti}{1996}]{alesina_1996} \textsc{Alesina, A. and Perotti, R.} (1996). Income Distribution, Political Instability, and Investment.
\textit{European Economic Review} \textbf{40}(6) 1203--1228.

\bibitem[\protect\citeauthoryear{Amemiya}{1978}]{amemiya_1978} \textsc{Amemiya, T.} (1978). A Note on a Random Coefficients Model.
\textit{International Economic Review} \textbf{19}(3) 793--796.

\bibitem[\protect\citeauthoryear{Arellano}{2003}]{arellano_2003b} \textsc{Arellano, M.}(2003) Modelling Optimal Instrumental Variables for Dynamic Panel Data Models. CEMFI, Madrid.

\bibitem[\protect\citeauthoryear{Arellano and Bond}{1991}]{arellano_1991} \textsc{Arellano, M. and Bond, S.} (1991). Some Tests of Specification for Panel Data: Monte Carlo Evidence and an Application to Employment Equations. \textit{The Review of Economic Studies} \textbf{58}(2) 277--297.

\bibitem[\protect\citeauthoryear{Azariadis}{1996}]{azariadis_1996} \textsc{Azariadis, C.} (1996). The Economics of Poverty Traps. Part One: Complete Markets.
\textit{Journal of Economic Growth} \textbf{1} 449--486.

\bibitem[\protect\citeauthoryear{Bandyopadhyay and Tang}{2011}]{bandyopadhyay_2010} \textsc{Bandyopadhyay, D and Tang, X.} (2011). Understanding the Economic Dynamics Behind Growth and Inequality Relationships.
\textit{Journal of Macroeconomics} \textbf{33} 14--32. 

\bibitem[\protect\citeauthoryear{Banerjee}{2000}]{banerjee_2000} \textsc{Banerjee, A.} (2000). The Two Poverties.
\textit{Massachusetts Institute of Technology,
Department of Economics WP} \textbf{1}(16).

\bibitem[\protect\citeauthoryear{Banerjee and Duflo}{2003}]{banerjee_2003} \textsc{Banerjee, A. and Duflo, E.} (2003). Inequality and Growth: What Can the Data Say?
\textit{Journal of Economic Growth} \textbf{8}(3) 267-299.

\bibitem[\protect\citeauthoryear{Banerjee and Newman}{1993}]{banerjee_1993} \textsc{Banerjee, A. V. and Newman, A. F.} (1993). Occupational Choice and the Process of Development.
\textit{Journal of Political Economy} \textbf{101}(2) 274--298.

\bibitem[\protect\citeauthoryear{Barro and Lee}{2010}]{barro_2010} \textsc{Barro, R. J. and Lee, J. W.} (2010). A New Data Set pf Educational Attainment in the World, 1950 -– 2010.
National Bureau of Economic Research, Working Paper 15902.

\bibitem[\protect\citeauthoryear{Benabou}{1996}]{benabou_1996a} \textsc{Benabou, R.} (1996). Heterogeneity, Stratification, and Growth: Macroeconomic Implications of Community Structure and School Finance.
\textit{American Economic Review} \textbf{86}(3) 584-609.

\bibitem[\protect\citeauthoryear{Bergh}{2010}]{bergh_2010} \textsc{Bergh, A. and Karlsson, M.} (2010). Government size and growth: Accounting for economic freedom and globalization.

\bibitem[\protect\citeauthoryear{Birdsall, Graham and Pettinato}{2000}]{birdsall_2000} \textsc{Birdsall, N., Graham, C. and Pettinato, S.} (2000). Stuck in the Tunnel: Is Globalization Muddling the Middle Class?
\textit{Center on Social and Economic Dynamics Working Paper 14}.

\bibitem[\protect\citeauthoryear{Blundell and Bond}{1998}]{blundell_1998} \textsc{Blundell, R. and Bond, S.} (1998). Initial Conditions and Moment Restrictions in Dynamic Panel Data Models. \textit{Journal of Econometrics} \textbf{87}(1) 115--143.

\bibitem[\protect\citeauthoryear{Bond, Hoeffler and Temple}{2001}]{bond_2001} \textsc{Bond, S. and Hoeffler, A. and Temple, J.}(2001). GMM Estimation of Empirical Growth Models. Discussion Paper No 01/525

\bibitem[\protect\citeauthoryear{Bourguignon}{1998}]{bourguignon_1998} \textsc{Bourguignon, F.} (1998). \'Equit\'e Et Croissance \'Economique: une nouvelle analyse?
\textit{Revue Fran\c{c}aise D'\'economie} \textbf{13}(3) 25--84.

\bibitem[\protect\citeauthoryear{Bourguignon}{2004}]{bourguignon_2004} \textsc{Bourguignon, F.} (2004). The Poverty-Growth-Inequality Triangle.
\textit{Indian Council for Research on International Economic Relations, New Delhi Working Papers}

\bibitem[\protect\citeauthoryear{Bowsher}{2002}]{bowsher_2002} \textsc{Bowsher, C.G.} (2002). On testing Overidentifying Restrictions in Dynamic Panel Data Models. \textit{Economics Letters}
\textbf{77} 211--220.


\bibitem[\protect\citeauthoryear{Bowles, Durlauf and Hoff}{2006}]{bowles_2006} \textsc{Bowles, S., Durlauf, S. N. and Hoff, K.} (2006). Poverty Traps.
\textit{Princeton University Press}


\bibitem[\protect\citeauthoryear{Caselli, Esquivel and Lefort}{1996}]{caselli_1996} \textsc{Caselli, F., Esquivel, G. and Lefort, F.} (1996). Reopening the Convergence Debate: A New Look at Cross-Country Growth Empirics. \textit{Journal of Economic Growth} \textbf{1}(3) 363--389.

\bibitem[\protect\citeauthoryear{Clemens, Radelet and Bhavnani}{2004}]{clemens_2004} \textsc{Clemens, M. A., Radelet, S. and Bhavnani, R. R.} (2004). Counting chickens when they hatch: The short term effect of aid on growth. \textit{Center for Global Development working paper} (44).

\bibitem[\protect\citeauthoryear{Cunha and Heckman}{2007}]{cunha_2007} \textsc{Cunha, F. and Heckman, J.} (2007). The Technology of Skill Formation.
\textit{American Economic Review} \textbf{97}(2) 31-47.

\bibitem[\protect\citeauthoryear{Chun, Hasan, Rahman and Uluba\c{s}o\u{g}lu}{2016}]{chun_2016} \textsc{Chun, N., Hasan, R. and Rahman, M. H. and Uluba\c{s}o\u{g}lu, M. A.} (20016). The Role of Middle Class in Economic Development: What Do Cross-Country Data Show?
\textit{Review of Development Economics}. forthcoming

\bibitem[\protect\citeauthoryear{Doepke and Zilibotti}{2005}]{doepke_2005} \textsc{Doepke, M. and Zilibotti, F.} (2005). Social Class and the Spirit of Capitalism.
\textit{Journal of The European Economic Association} \textbf{3}(2-3) 516-524.

\bibitem[\protect\citeauthoryear{Easterly}{2001}]{easterly_2001} \textsc{Easterly, W.} (2001). The Middle Class Consensus and Economic Development.
\textit{Journal of Economic Growth} \textbf{6}(4) 317--335.

\bibitem[\protect\citeauthoryear{Deininger and Squire}{1999}]{deininger_1996} \textsc{Deininger, K. and Squire, L.} (1996). A New Data Set Measuring Income Inequality.
\textit{The World Bank Economic Review} \textbf{10}(3) 565--591. 


\bibitem[\protect\citeauthoryear{Forbes}{2000}]{forbes_2000} \textsc{Forbes, K. J.} (2000). A Reassessment of the Relationship between Inequality and Growth.
\textit{The American Economic Review} \textbf{90}(4) 869--887.

\bibitem[\protect\citeauthoryear{Galor and Zeira}{1993}]{galor_1993} \textsc{Galor, O. and Zeira, J.} (1993). Income Distribution and Macroeconomics.
\textit{The Review of Economic Studies} \textbf{60}(1) 35--52.

\bibitem[\protect\citeauthoryear{Galor and Tsiddon}{1997}]{galor_1997a} \textsc{Galor, O. and Tsiddon, D.} (1997). The Distribution of Human Capital and Economic Growth.
\textit{Journal of Economic Growth} \textbf{2}(1) 93--124.

\bibitem[\protect\citeauthoryear{Galor and Tsiddon}{1997}]{galor_1997b} \textsc{Galor, O. and Tsiddon, D.} (1997). Technological Progress, Mobility, and Economic Growth.
\textit{American Economic Review} \textbf{87}(3) 363--382.

\bibitem[\protect\citeauthoryear{Ghatak}{2015}]{ghatak_2015} \textsc{Ghatak, M.} (2015). Theories of poverty traps and anti-poverty policies.
\textit{The World Bank Economic Review} \textbf{29}(suppl 1) 77--105.

\bibitem[\protect\citeauthoryear{Go, Nikitin, Wang and Zou}{2007}]{go_2007} \textsc{Go, D., Nikitin, D., Wang, X. and Zou, H.} (2007). Poverty and Inequality in Sub-Saharan Africa: Literature Survey and Empirical Assessment.
\textit{Annals Of Economics And Finance} \textbf{8}(2) 251--304.


\bibitem[\protect\citeauthoryear{Goldfarb and Idnani}{1982}]{goldfarb_1982} \textsc{Goldfarb, D. and Idnani, A.} (1982). Dual and Primal-Dual Methods for Solving Strictly Convex Quadratic Programs.
\textit{Numerical Analysis} 226--239. 


\bibitem[\protect\citeauthoryear{Goldfarb and Idnani}{1983}]{goldfarb_1983} \textsc{Goldfarb, D. and Idnani, A.} (1983). A Numerically Stable Dual Method For Solving Strictly Convex Quadratic Programs.
\textit{Mathematical Programming} \textbf{27}(1) 1--33.


\bibitem[\protect\citeauthoryear{Hahn and Kuersteiner}{2002}]{hahn_2002} \textsc{Hahn, J. and Kuersteiner, G.} (2002). Asymptotically Unbiased Inference for a Dynamic Panel Model with Fixed Effects When Both n and T Are Large. \textit{Econometrica} \textbf{70}(4) 1639--1657.

\bibitem[\protect\citeauthoryear{Hansen}{1982}]{hansen_1982} \textsc{Hansen, L. P.} (1982). Large Sample Properties of Generalized Method of Moments Estimators. \textit{Econometrica}
\textbf{50}(4) 1029--1054.

\bibitem[\protect\citeauthoryear{Halter, Oechslin and Zweim\"uller}{2014}]{halter_2014} \textsc{Halter, D., Oechslin, M. and Zweim\"uller, J.} (2014). Inequality and growth: the neglected time dimension. \textit{Journal of Economic Growth}
\textbf{19}(1) 81--104.

\bibitem[\protect\citeauthoryear{Heidenreich, Schindler and Sperlich}{2010}]{heidenreich_2010} \textsc{Heidenreich, N.B. and Schindler, A. and Sperlich, S.}(2010). Bandwidth Selection Methods for Kernel Density Estimation - A Review of Performance, SSRN-id1726428. 

\bibitem[\protect\citeauthoryear{Heston, Summers and Aten}{2009}]{heston_2009} \textsc{Heston, A. and Summers, R. and Aten, B.} (2009). Penn World Table Version 6.3.
\textit{Center for International Comparisons of Production, Income and Prices at the University of Pennsylvania}.

\bibitem[\protect\citeauthoryear{Hodrick and Prescott}{1997}]{hodrick_1997} \textsc{Hodrick, R. J. and Prescott, E. C.} (1997). Postwar U.S. Business Cycles: An Empirical Investigation.
\textit{Journal of Money, Credit and Banking} \textbf{29}(1) 1--16.

\bibitem[\protect\citeauthoryear{Hoeffler}{2002}]{hoeffler_2002} \textsc{Hoeffler, A.E.} (2002). The augmented Solow model and the African growth debate.
\textit{Oxford Bulletin of Economics and Statistics} \textbf{64}(2) 135--158.

\bibitem[\protect\citeauthoryear{Islam}{1995}]{islam_1995} \textsc{Islam, N.} (1995). Growth Empirics: A Panel Data Approach.
\textit{The Quarterly Journal of Economics} \textbf{110}(4) 1127--70. 

\bibitem[\protect\citeauthoryear{Jalan and Ravallion}{2004}]{jalan_2004} \textsc{Jalan, J. and Ravallion, M.} (2004). Insurance Against Poverty (Chapter 5).
\textit{Oxford University Press}.

\bibitem[\protect\citeauthoryear{Kaldor}{1956}]{kaldor_1956} \textsc{Kaldor, N.} (1956). Alternative Theories of Distribution.
\textit{The Review of Economic Studies} \textbf{23}(2) 83--100.

\bibitem[\protect\citeauthoryear{Kao and Chiang}{2000}]{kao_2000} \textsc{Kao, C. and Chiang, M. H.} (2000). On The Estimation And Inference Of A Cointegrated Regression In Panel Data.
\textit{Advances in Econometrics} \textbf{15} 179--222.

\bibitem[\protect\citeauthoryear{Kendall}{1954}]{kendall_1954} \textsc{Kendall, M.G.} (1954). Note on the Bias in the Estimation of Autocorrelations. \textit{Biometrika} \textbf{41} 403--404.

\bibitem[\protect\citeauthoryear{Kiviet}{1995}]{kiviet_1995} \textsc{Kiviet, J.F.} (1995). On Bias, Inconsistency, and Efficiency of Various Estimators in Dynamic Panel Data Models. \textit{Journal of Econometrics} \textbf{68} 53--78.

\bibitem[\protect\citeauthoryear{Koehler, Sperlich and Vortmeyer}{2011}]{koehler_2011} \textsc{Max Koehler, M., Sperlich, S. and Vortmeyer, J.}(2011). The Africa-Dummy in Growth Regressions. Courant Research Centre "Poverty, Equity and Growth in Developing and Transition Countries: Statistical Methods and Empirical Analysis", Discussion Paper No 94. 


\bibitem[\protect\citeauthoryear{Kraay}{2006}]{kraay_2006} \textsc{Kraay, A.} (2006). When is Growth Pro-Poor? Evidence from a Panel of Countries.
\textit{Journal of Development Economics} \textbf{80} 198--227.

\bibitem[\protect\citeauthoryear{Kurita and Kurosaki}{2011}]{kurita_2011} \textsc{Kurita, K. and Kurosaki, T.} (2011). The Dynamics of Growth, Poverty and Inequality: A Panel Analysis of Regional Data from Thailand and the Philippines.
\textit{Asian Economic Journal} \textbf{25}(1) 3--33.

\bibitem[\protect\citeauthoryear{Kuznets}{1955}]{kuznets_1955} \textsc{Kuznets, S} (1955). Economic Growth and Income Inequality.
\textit{The American Economic Review} \textbf{45}(1) 1--28.

\bibitem[\protect\citeauthoryear{Landes}{1999}]{landes_1999} \textsc{Landes, D.} (1999). The Wealth and Poverty of Nations: Why Some Are So Rich and Some So Poor.
\textit{Norton} \textbf{8}(3).

\bibitem[\protect\citeauthoryear{Lokshin and Ravallion}{2004}]{lokshin_2004} \textsc{Lokshin, M. and Ravallion, M.} (2004). Household Income Dynamics in Two Transition Economies.
\textit{Studies in Nonlinear Dynamics and Econometrics} \textbf{8}(3).

\bibitem[\protect\citeauthoryear{Lopez and Serv\'en}{2009}]{lopez_2009} \textsc{Lopez, H. and Serv\'en, L.} (2009). Too Poor to Grow.
\textit{Policy Research Working Paper 5012}.

\bibitem[\protect\citeauthoryear{Malinen}{2013}]{malinen_2013} \textsc{Malinen, T.} (2013). Inequality and growth: Another look with a new measure and method.
\textit{Journal of International Development} \textbf{25}(1) 122--138.

\bibitem[\protect\citeauthoryear{Mankiw, Romer and Weil}{1992}]{mankiw_1992} \textsc{Mankiw, N. G., Romer, D. and Weil, D. N.} (1992). A Contribution to the Empirics of Economic Growth.
\textit{The Quarterly Journal of Economics} \textbf{107}(2) 407--437.

\bibitem[\protect\citeauthoryear{Marrero and Rodr\'{i}guez}{2013}]{marrero_2013} \textsc{Marrero, G. A. and Rodr\'{i}guez, J. G.} (1992). Inequality of opportunity and growth.
\textit{Journal of Development Economics} \textbf{104} 107--122.

\bibitem[\protect\citeauthoryear{McKenzie and Woodruff}{2006}]{mckenzie_2006} \textsc{McKenzie, D.J. and Woodruff, C.} (2006). Do Entry Costs Provide an Empirical Basis for Poverty Traps? Evidence from Mexican Microenterprises.
\textit{Economic Development and Cultural Change} \textbf{55}(1) 3--42.

\bibitem[\protect\citeauthoryear{Mesnard and Ravallion}{2006}]{mesnard_2006} \textsc{Mesnard, A. and Ravallion, M.} (2006). The Wealth Effect on New Business Startups in a Developing Economy.
\textit{Economica} \textbf{73} 367--392.

\bibitem[\protect\citeauthoryear{Murphy, Schleifer and Vishny}{1989}]{murphy_1989} \textsc{Murphy, K. M. and Schleifer, A. and Vishny, R. W.} (1989). Industrialization and the Big Push.
\textit{Journal of Political Economy} \textbf{97}(5) 1003-1026.

\bibitem[\protect\citeauthoryear{Nickell}{1981}]{nickell_1981} \textsc{Nickell, S.} (1981). Biases in Dynamic Models with Fixed Effects. \textit{Econometrica} \textbf{49}(6) 1417--26.

\bibitem[\protect\citeauthoryear{Orcutt and Irwin}{1948}]{orcutt_1948} \textsc{Orcutt, G. H. and Irwin, J. O.} (1948). A Study of the Autoregressive Nature of the time series Used for Tinbergen's Model of the Economic System of the United States. \textit{Journal of Royal Statistical Society} \textbf{10}(1) 1--53.

\bibitem[\protect\citeauthoryear{Park, Mammen, Lee and Lee}{2013}]{PBMLL2013} 
\textsc{Park, B. U., Mammen, E., Lee, Y. K. and Lee, E. R.} (2013). Varying Coefficient Regression Models: A Review and New Developments. \textit{International Statistical Review} \textbf{83}(1) 36--64.

\bibitem[\protect\citeauthoryear{Piketty}{1993}]{piketty_1993} \textsc{Piketty, T.} (1993). Imperfect Capital Markets and the Persistence of Initial Wealth Inequalities.
\textit{London School of Economics Suntory Toyota Centre for Economics and Related Disciplines Working Paper} \textbf{TE}(92) 255. 

\bibitem[\protect\citeauthoryear{Persson and Tabelline}{1990}]{persson_1990} \textsc{Persson, T. and Tabelline, G.} (1990). Politico-Economic Equilibrium Growth: Theory and Evidence.
\textit{mimeo} 

\bibitem[\protect\citeauthoryear{Persson and Tabellini}{1994}]{persson_1994} \textsc{Persson, T. and Tabellini, G.} (1994). Is Inequality Harmful for Growth?
\textit{American Economic Review} \textbf{84}(3) 600--621.

\bibitem[\protect\citeauthoryear{Perry, Lopez and Maloney}{2006}]{perry_2006} \textsc{Perry, G.E. and Lopez, J.H. and Maloney, W.F.} (2006). Poverty Reduction and Growth: Virtuous and Vicious Circles.
\textit{The World Bank}.

\bibitem[\protect\citeauthoryear{Phillips and Sul}{2007}]{phillips_2007} \textsc{Phillips, P. C. B. and Sul, D.} (2007). Bias in dynamic panel estimation with fixed effects, incidental trends and cross section dependence. \textit{Journal of Econometrics} \textbf{137} 162--188.


\bibitem[\protect\citeauthoryear{Phillips and Moon}{1999}]{phillips_1999} \textsc{Phillips, P. C. B. and Moon, H.} (1999). Linear Regression Limit Theory for Nonstationary Panel Data. \textit{Econometrica} \textbf{67} 1057--1111.

\bibitem[\protect\citeauthoryear{Pritchett}{1996}]{pritchett_1996} \textsc{Pritchett, L.} (1996). Where Has All the Education Gone. The World Bank, Policy Research Working Paper 1581.

\bibitem[\protect\citeauthoryear{Rajan and Subramanian}{2008}]{rajan_2008} \textsc{Rajan, R. G. and Subramanian, A.} (2008). Aid and growth: What does the cross-country evidence really show?
\textit{The Review of economics and Statistics} \textbf{90}(4), 643-665.

\bibitem[\protect\citeauthoryear{Ravallion}{2010}]{ravallion_2010} \textsc{Ravallion, M.} (2010). Why Dont We See Poverty Convergence?
\textit{Development Research Group, World Bank, Working Paper} \textbf{4974}

\bibitem[\protect\citeauthoryear{Ravallion and Chen}{2003}]{ravallion_2003} \textsc{Ravallion, M. and Chen, S.} (20003). Measuring Pro-Poor Growth
\textit{Economic Letters} \textbf{78}(1) 93--99. 

\bibitem[\protect\citeauthoryear{Rodr\'{\i}guez-Po\'{o} and Sober\'{o}n}{2015}]{RodPooSoberon2015}
Rodr\'{\i}guez-Po\'{o}, J. and A. Sober\'{o}n (2015). Nonparametric estimation of fixed effects panel data varying coefficient models \textit{Journal of Multivariate Analysis} \textbf{133} 95--122.


\bibitem[\protect\citeauthoryear{Rodrik}{1998}]{rodrik_1998} \textsc{Rodrik, D.} (1998). Where Did All the Growth Go? External Shocks, Social Conflict and Growth Collapses.
\textit{Centre for Economic Policy Research Discussion Paper} \textbf{1789}.

\bibitem[\protect\citeauthoryear{Roodman}{2006}]{roodman_2006} \textsc{Roodman, D.}(2006). How to Do xtabond2: An Introduction to Difference and System GMM in Stata. Center for Global Development, Working Paper Number 103.

\bibitem[\protect\citeauthoryear{Roodman}{2009}]{roodman_2009} \textsc{Roodman, D.} (2009). A Note on the Theme of Too Many Instruments. \textit{Oxford Bulletin of Economics and Statistics} \textbf{71}(1) 135--158.

\bibitem[\protect\citeauthoryear{Sachs, McArthur, Schmidt-Traub and Kruk}{2004}]{sachs_2004} \textsc{Sachs, J. D. and McArthur, J. W. and Schmidt-Traub, G. and Kruk, M.} (2004). Ending Africa's Poverty Trap.
\textit{Brookings Papers on Economic Activity} \textbf{2004}(1) 117--216.

\bibitem[\protect\citeauthoryear{Saint-Paul and Verdier}{1993}]{saint-paul_1993} \textsc{Saint-Paul, G. and Verdier, T.} (1993). Education, Democracy, and Growth.
\textit{Journal of Development Economics} \textbf{42}(2) 399--407.

\bibitem[\protect\citeauthoryear{Sala-i-Martin}{2006}]{sala-i-martin_2006} \textsc{Sala-i-Martin, X.} (2006). The World Distribution of Income: Falling Poverty and Convergence, Period.
\textit{Quarterly Journal of Economics} \textbf{121}(2) 351--397. 

\bibitem[\protect\citeauthoryear{Singh, Nagar, Choudhry and Baldev}{1976}]{singh_1976} \textsc{Singh, B., Nagar, A. L., Choudhry, N. K. and Baldev,R.} (1993). On the Estimation of Structural Change: A Generalization of the Random Coefficients Regression Model.
\textit{International Economic Review} \textbf{17}(2) 340--361.

\bibitem[\protect\citeauthoryear{Sridharan}{2004}]{sridharan_2004} \textsc{Sridharan, E.} (2004). The Growth and Sectoral Composition of India's Middle Class: Its Impact on the Politics of Economic Liberalization.
\textit{India Review} \textbf{3}(4) 405--428.

\bibitem[\protect\citeauthoryear{Tapia}{2015}]{tapia_2015} \textsc{Tauchen, G.} (1986). Inequality and poverty in a developing economy: Evidence from regional data (Spain, 1860-1930). \textit{EHES Working Papers In Economic History} \textbf{78}.

\bibitem[\protect\citeauthoryear{Tauchen}{1986}]{tauchen_1986} \textsc{Tauchen, G.} (1986). Statistical Properties of Generalized Method-of-Moments Estimators of Structural Parameters Obtained from Financial Market Data. \textit{Journal of Business and Economic Statistics} \textbf{4}(4) 397--416.

\bibitem[\protect\citeauthoryear{The World Bank}{2009}]{worldbank_2009} \textsc{The World Bank} (2005). World Development Indicators (producer and distributor). http://data.worldbank.org/data-catalog/world-development-indicators

\bibitem[\protect\citeauthoryear{Wachter}{1970}]{wachter_1970} \textsc{Wachter, M. L.} (1970). Relative Wage Equations for U.S. Manufacturing Industries 1947-1967. \textit{The Review of Economics and Statistics} \textbf{52}(4) 405--410.

\bibitem[\protect\citeauthoryear{Windmeijer}{2005}]{windmeijer_2005} \textsc{Windmeijer, F.} (2005). A Finite Sample Correction for the Variance of Linear Efficient Two-Step GMM Estimators. \textit{Journal of Econometrics} \textbf{126}(1) 25--51.


\bibitem[\protect\citeauthoryear{Xia}{2010}]{xia_2010} \textsc{Xia, B.} (2010). Status, Inequality and Intertemporal Choice.
\textit{The B.E. Journal of Theoretical Economics} \textbf{10}(1).

\bibitem[\protect\citeauthoryear{Ziliak}{1997}]{ziliak_1997} \textsc{Ziliak, J. P.} (1997). Efficient Estimation with Panel Data when Instruments Are Predetermined: An Empirical Comparison of Moment-Condition Estimators. \textit{Journal of Business and Economic Statistics} \textbf{15}(4) 419--431.

\end{thebibliography}
\end{document}